\documentclass[sn-mathphys-num]{sn-jnl}

\usepackage{amssymb}
\usepackage{amsmath}
\usepackage{pgfplots}
\usepackage{enumitem}
\usepackage{graphicx} 
\usepackage{hyperref} 
\usepackage{natbib} 
\usepackage{lineno} 
\usepackage{geometry} 
\usepackage{multirow}
\usepackage{multicol}
\usepackage{longtable}
\usepackage{algorithm}
\usepackage{nicematrix}
\usepackage{booktabs}
 \providecommand{\floor}[1]{\left \lfloor #1 \right \rfloor }
 \usepackage{float}
 \usepackage[font=footnotesize,caption=false]{subfig}
 \usepackage{lscape}
 \usepackage{etoolbox}
 \usepackage{framed}
 \usepackage{xcolor}[2007/01/21]
 \usepackage{makecell}
 
 \usepackage{rotating}
 \usepackage{hhline}
\usepackage{graphicx}%
\usepackage{multirow}%
\usepackage{amsmath,amssymb,amsfonts}%
\usepackage{amsthm}%
\usepackage{mathrsfs}%
\usepackage[title]{appendix}%
\usepackage{xcolor}%
\usepackage{textcomp}%
\usepackage{manyfoot}%
\usepackage{booktabs}%
\usepackage{algorithm}%
\usepackage{algorithmicx}%
\usepackage{algpseudocode}%
\usepackage{listings}%
\usepackage{tabularx}

\theoremstyle{thmstyleone}%
\theoremstyle{thmstyletwo}%

\theoremstyle{thmstylethree}%
\newtheorem{definition}{Definition}%

\begin{document}

\title[Article Title]{G{\"o}del Number based Clustering Algorithm with Decimal First Degree Cellular Automata}

\author{\fnm{Vicky} \sur{Vikrant}}\email{vicky.vikrant8396@gmail.com}

\author{\fnm{Narodia} \sur{Parth P}}\email{ppnarodiya27@gmail.com}

\author{\fnm{Kamalika} \sur{Bhattacharjee}}\email{kamalika.it@gmail.com}

\affil{\orgdiv{Department of Computer Science and Engineering}, \orgname{National Institute of Technology}, \orgaddress{\street{Thuvakudi}, \city{Tiruchirappalli}, \postcode{620015}, \state{Tamil Nadu}, \country{India}}}

%


\abstract{In this paper, a decimal first degree cellular automata (FDCA) based clustering algorithm is proposed where clusters are created based on \emph{reachability}. \emph{Cyclic spaces} are created and configurations which are in the same cycle are treated as the same cluster. Here, real-life data objects are encoded into decimal strings using G{\"o}del number based encoding. The benefits of the scheme is, it reduces the encoded string length while maintaining the features properties. Candidate CA rules are identified based on some theoretical criteria such as \emph{self-replication} and \emph{information flow}. An iterative algorithm is developed to generate the desired number of clusters over three stages. The results of the clustering are evaluated based on benchmark clustering metrics such as Silhouette score, Davis Bouldin, Calinski Harabasz and Dunn Index. In comparison with the existing state-of-the-art clustering algorithms, our proposed algorithm gives better performance.}




\maketitle
%
%







\section{Introduction}
\label{sec:Introduction }
Clustering with respect to bijectivity refers to the process of maintaining a one-to-one relationship between data points and clusters in a clustering method. A one-to-one function divides the objects in the domain into discrete, non-intersecting groups, each representing a single cluster. There exist many state-of-the-art algorithms, such as K-Means~\cite{hartigan1979algorithm}, Hierarchical~\cite{Hierarchical}, DBSCAN~\cite{ester1996density} etc for clustering. Clustering measures and benchmarks are used to assess quality and performance of the generated clusters. 
The problem of clustering data points is a common challenge in unsupervised learning, providing valuable insights into datasets. However, existing algorithms often lack for large and real-life datasets~\cite{mukherjee2021clustering, mukherjee2021reversible}, producing clusters with no significant interpretations. The difficulty increases as the data dimensionality increases, as several features complicate the clustering process and computational needs. While dimensionality reduction techniques can be used, they frequently need to be better because they may lose key features and fail to build accurate clusters. 

For example, K-Means performs poorly on huge datasets when no strategies are utilized. Principal Component Analysis can be used to reduce the dimensions. However, this is not advised for huge datasets, where each feature is equally important because they can result in meaningless clusters indicated by the low scores in the benchmark metrics. Recently, reversible cellular automaton (CA) have been proposed as a natural technique of clustering.
In a reversible CA, the one-to-one global transition function divides the configurations into a set of arrangements that can be \emph{reached} from one another, establishing a cycle that defines a cluster. As a result, arrangements that can be evolved into one other are regarded as \emph{close} and belong to the same cluster, but those that cannot be reached into each other belong to separate clusters. This \emph{reachability} metric is necessary for building groups with reversible CAs. However, all these CA based algorithm work for binary numbers. But real-life datasets usually are not in binary. So, encoding techniques are used to convert real datasets into binary format to do clustering with binary cellular automata (CAs). To deal with high-dimensional datasets in a computationally effective way, the concept of vertical splitting was introduced~\cite{abhishek2023cellular}.
But, although these algorithm perform at par with the other state-of-the-art algorithms, all of them, including the state-of-the-art algorithms, fail to achieve the optimal clustering scores. The reason may be, in the process of encoding into binary by the chosen encoding techniques, or otherwise, the properties of the features are lost making way for the low clustering scores.

There are many possible options to address this problem, such as direct conversion, hashing and existing encoding techniques. These work well for a limited number of characteristics, but for larger datasets like Big Data, the encoded configuration length can be significantly longer. For example, in a high-dimensional datasets, there may be more than 50 features with each feature having, say, 3 digit representation, then if the direct concatenation is used to construct a string from one row by merging all feature values, the string length may reach 150 digits which is computationally very costly for clustering. Similarly, if existing hashing techniques are used, the properties of the features are lost giving bad clusters.
In this situation, we need some encoding technique that compress the dataset to reduce the size of the dataset without compromising on the properties of the features.  

To overcome these restrictions in this paper, a novel algorithm based on decimal reversible cellular automata is developed. This algorithm outperforms existing approaches and effectively manages real-world data points. The technique works by encoding data points having any number of features into decimal strings, which serve as inputs to reversible cellular automata. G{\"o}del numbering is used for this encoding, which significantly reduces string length while maintaining the properties and features as given in the original dataset (The concept of G{\"o}del number based encoding for usage in clustering was introduced in Ref.~\cite{narodia2023godel}; in this paper we improve and extend that concept with application to cellular automata based clustering.). This minimizes computational complexity while improving the quality of clustering results. As this encoding creates decimal numbers, to use this encoding over a numerical dataset, we need decimal cellular automata.
However, number of possible decimal CAs over even the nearest (three) neighborhood dependency is gigantic ($10^{10^{3}}$). This makes the set impossible to theoretically analyze and experiment on to find the set of good CAs for clustering. For this reason, we choose the decimal first degree cellular automata (FDCA) in our work.

The first degree cellular automata provide a framework for capturing local interactions and updating states using basic rules, which can be used in clustering tasks. However, the effectiveness of first degree CA in clustering might vary based on the dataset, problem complexity, and the precise rule definitions and parameter settings employed. Experimentation and customization are frequently required to adapt the method to the unique dataset and clustering challenge at hand. In our work, we are going to use G{\"o}del number based encoding and decimal first degree cellular automata for clustering and select rules with the help of theoretical properties such as chaotic parameters and analysis on cyclic spaces. Nevertheless, only a small subset of these decimal CAs are studied, improved outcomes are possible over other CAs.

\section{Preliminaries}
\label{sec: Preliminaries}
This section provides a concise overview of the technical terms used in the paper.

\subsection{Cellular Automata Basics}
\label{sec:Preliminaries}
The application of cellular automata in clustering refers to the application of simple discrete dynamical systems with local interactions involving immediate (or, almost immediate) neighbors to solve clustering problem. This work considers $3$-neighborhood $10$-state finite CAs under null boundary condition such that the lattice $\mathcal{L}  = \mathbb{Z}/n\mathbb{Z}$ where $n$ is the number of cells. The state set $S = \{0,1,\cdots, 9\}$ and each cell depends on its left cell, itself and its right cell. So, the rule is $R: S^3 \rightarrow S$. Let the arguments of $R$ be a triplet $\langle x,y,z\rangle$ where $x,y,\&~ z$ represent the left, self and right neighbors of a cell respectively. This triplet is named as \emph{Rule Min Term} or simply an \emph{RMT} and represented by $r=x \times 10^2 + y \times 10 + z$. So, a rule has $1000$ number of RMTs. There exists several grouping of the RMTs:

\begin{definition}[Equivalent RMTs]
	\label{equi}
	A set of $10$ RMTs $r_0, r_1,\cdots, r_{9}$ of a $10$-state CA rule are said to be equivalent to each other if, for any $r_i, r_j$, where $r_i \ne r_j$, $r_i\equiv r_j\pmod{10^2}$~\cite{BHATTACHARJEEdecimalCA}.  
\end{definition}

\begin{definition}[Sibling RMTs]
	\label{sibl}
	A set of $10$ RMTs $s_0, s_1, \cdots, s_{9}$ of a $10$-state CA rule are said to be sibling to each other if, for any $s_i,s_j$, where $s_i \ne s_j$, $\floor{\frac{s_i}{10}} = \floor{\frac{s_j}{10}}$~\cite{BHATTACHARJEEdecimalCA}.
\end{definition} 
For all $i$, $0\le i\le 99$, two sets can be formed -- $Equi_i=\{r~|~ r = k\times 10^2+i, 0\le k\le 9\}$ and $Sibl_i=\{s~|~ s= 10\times i+k, 0\le k \le 9\}$. Similarly, each RMT has a corresponding $L$-set and $R$-set.

\begin{definition}[L-Set]
	For any RMT $r=x \times 10^2 + y \times 10 + z$, $L$-set($r$) = $\{s~|~s=10^2\times x + 10\times y'+z'$ for all $y'\ne y$ and $z'\ne z, y',z'\in S\}$~\cite{Supreeti_2018_chaos}.
\end{definition}

\begin{definition}[R-Set]
	For any RMT $r=x \times 10^2 + y \times 10 + z$, $R$-set(r) =  $\{s~|~s=10^2\times x' + 10\times y'+z$ for all $x'\ne x$ and $y'\ne y, x'y'\in S\}$~\cite{Supreeti_2018_chaos}.
\end{definition}

\begin{table}[!h]
	\centering
	\vspace{-1em}
	\caption{Relations among Equivalent and Sibling RMTs for d = 10}
	\label{Equivalent and Sibling RMTs}
		\begin{tabular}{|l|l|l|}
			\hline
			\textbf{\#Set} & \textbf{Sibling RMTs}           & \textbf{Equivalent RMTs}                 \\ \hline
			0             & \{0,1,2,3,4,5,6,7,8,9\}           & \{0,100,200,300,400,500,600,700,800,900\}  \\ \hline
			1             & \{10,11,12,13,14,15,16,17,18,19\} & \{1,101,201,301,401,501,601,701,801,901\}  \\ \hline
			2             & \{20,21,22,23,24,25,26,27,28,29\} & \{2,102,202,302,402,502,602,702,802,902\}  \\ \hline
			\begin{tabular}[c]{@{}l@{}}.\\ .\\ .\end{tabular} & \begin{tabular}[c]{@{}l@{}}.\\ .\\ .\end{tabular} & \begin{tabular}[c]{@{}l@{}}.\\ .\\ .\end{tabular} \\ \hline
			99            & \{990,991,992,993,994,995,996,997,998,999\} & \{99,199,299,399,499,599,699,799,899,999\} \\ \hline
	\end{tabular}
	\vspace{-1em}
\end{table}

\begin{table}[!h]
	\centering
	
	\caption{Sample L-Set}
	\label{L-Set for every RMT}
	\scriptsize
		\begin{tabular}{|ccc|}
			\hline
			\multicolumn{1}{|c|}{\textbf{Set}}       & \textbf{L-set(r)}         &            \\ \hline
			\multicolumn{1}{|c|}{\textbf{L-set(0)}}  & \{011,012,013,$\cdots$,019, 021, 022,$\cdots$,029,$\cdots$,091,092,$\cdots$,099\} &\\ \hline 
			\multicolumn{1}{|c|}{\textbf{L-set(1)}}  & \{010,012,013,$\cdots$,019, 020,022,$\cdots$,029,$\cdots$,090,092,$\cdots$,099\} &  \\ \hline 
			\multicolumn{1}{|c|}{\textbf{\begin{tabular}[c]{@{}c@{}}$\vdots$\end{tabular}}} &
			\begin{tabular}[c]{@{}c@{}}$\vdots$\end{tabular} & \\ \hline
			
			\multicolumn{1}{|c|}{\textbf{L-set(99)}} & \{000,001,002,$\cdots$008, 010,011,$\cdots$,018,$\cdots$,080,081,$\cdots$ ,088\} & \\ \hline
			\multicolumn{1}{|c|}{\textbf{\begin{tabular}[c]{@{}c@{}}$\vdots$\end{tabular}}} &
			\begin{tabular}[c]{@{}c@{}}$\vdots$\end{tabular} &
		\\ \hline
	\end{tabular}
	\vspace{-1em}
	
\end{table}

\begin{table}[!h]
	\centering
	
	\caption{Sample R-Set}
	\label{R-Set for every RMT}
	\scriptsize
		\begin{tabular}{|cc|cc|}
			\hline
			
			\multicolumn{1}{|c|}{\textbf{Set}}       & \textbf{R-set(r)}         \\ \hline
			\multicolumn{1}{|c|}{\textbf{R-set(0)}}  & \{110,120,130,$\cdots$,190, 210,220,$\cdots$,290,$\cdots$,910,920,$\cdots$,990\} \\ \hline
			
			 \multicolumn{1}{|c|}{\textbf{R-set(1)}}  & \{111,121,131,$\cdots$, 191, 211,221,$\cdots$,291,$\cdots$,911,921,$\cdots$,991\} \\ \hline
		
			\multicolumn{1}{|c|}{\textbf{\begin{tabular}[c]{@{}c@{}}$\vdots$\end{tabular}}} &
			\begin{tabular}[c]{@{}c@{}}$\vdots$\end{tabular} \\ \hline
		     \multicolumn{1}{|c|}{\textbf{R-set(99)}} & \{109,119,129,189, 209,219,$\cdots$289,$\cdots$,909,919,$\cdots$,989\} \\ \hline
		
			\multicolumn{1}{|c|}{\textbf{\begin{tabular}[c]{@{}c@{}}$\vdots$\end{tabular}}} &
			\begin{tabular}[c]{@{}c@{}}$\vdots$\end{tabular} \\ \hline
		\end{tabular}
	\vspace{-1em}
	
\end{table}

Table~\ref{Equivalent and Sibling RMTs} depicts the equivalent and sibling RMT sets for such a decimal CA. Table~\ref{L-Set for every RMT} and Table~\ref{R-Set for every RMT} shows the $L$-set and $R$-sets for each of the RMTs in a $3$-neighborhood $10$-state CA.

\begin{definition}\label{Def:permutivity}
	Let $R: S^3 \rightarrow S$ be the local rule of a CA. Then, the CA is left-permutive (resp. right-permutive) if and only if for any two equivalent (resp. sibling) RMTs $r$ and $s$, $R[r]\ne R[s]$~\cite{BHATTACHARJEEdecimalCA}.
\end{definition}
Therefore, if, for all $i$, all RMTs of $Sibl_i$ (resp. $Equi_i$) have distinct next state values, the CA is a right-permutive (resp. left-permutive) CA. An RMT $r = x \times 10^2 + y \times 10 + z$ is said to be self-replicating, if $R(x,y,z)=y$, where $R$ is the rule of the CA~\cite{BHATTACHARJEEdecimalCA}. Self-replicating RMTs play an important rule in stabilizing the system.


\subsection{First Degree Cellular Automata}
\label{sec:PreliminariesFDCA}
Naturally, representing a rule with 1000 RMTs is often a tedious task. For that reason first degree cellular automata (FDCA) has been proposed which are very easy to represent and work with~\cite{bhattacharjee2023study}. These CA are a subset of the possible CAs for any state with three neighborhood condition. Formally, a decimal FDCA can be defined as follows:

\begin{definition}
	A decimal cellular automata rule $R: S^3 \rightarrow S$ is of first degree\ if the rule can be represented in the following form:
	\begin{equation}
		\label{degree1}
		R(x,y,z)= (c_0xyz + c_1xy +c_2xz + c_3yz +c_4x +c_5y +c_6 z +c_7) \pmod{10}
	\end{equation}
	Here, $10$ is the number of states of the CA, $c_i \in \mathbb Z_{10}=\{0, 1, \cdots, 9\}$ and $x,y,z$ are the three neighbors of each cell where $x,y,z \in S$~\cite{bhattacharjee2022first}. 
\end{definition}
For each rule, $c_i$ represents a constant value that is fixed for the rule. In Equation~\ref{degree1}, each of $x$, $y$, and $z$ is of degree one, so we refer to this CA rule as a \emph{rule of first degree} or simply a \emph{first degree CA}.
Any rule based on Equation~\ref{degree1} can be expressed solely by the constant values $\langle c_0, c_1, \cdots, c_7\rangle $; we refer to these values as the \emph{parameters} of the first degree CA. Our rule space is effectively reduced by selecting only CAs represented by these parameters. For example, when $d=2$, all $2^8 = 256$ rules are covered. However, for $10$, there are only $10^8$ first degree CAs, out of possible $10^{1000}$ number of $3$-neighborhood decimal CAs. Table~\ref{FDCA Eg} shows an example of decimal FDCA $R$ represented by the parameters along with the corresponding rule $R$ in the format of concatenating the values $R[r]$ for each of the $1000$ RMTs $r$ such that $999\ge r\ge 0$.
\begin{table}[hbtp]
	\centering
	\caption{A sample decimal FDCA rule}
	\label{FDCA Eg}
	\begin{tabular}{|c|c|}
		\hline
		\textbf{FDCA Parameter} &
		\textbf{Actual Rule} \\ \hline
		$\langle0,0,0,0,1,0,1,8\rangle$ &
		\begin{tabular}[c]{@{}c@{}}65432109876543210987654321098765432109876543210987654\\ 32109876543210987654321098765432109876543210987543210\\ 98765432109876543210987654321098765432109876543210987\\ 65432109876543210987654321098765432109876432109876543\\ 21098765432109876543210987654321098765432109876543210\\ 98765432109876543210987654321098765321098765432109876\\ 54321098765432109876543210987654321098765432109876543\\ 21098765432109876543210987654210987654321098765432109\\ 87654321098765432109876543210987654321098765432109876\\ 54321098765432109876543109876543210987654321098765432\\ 10987654321098765432109876543210987654321098765432109\\ 87654321098765432098765432109876543210987654321098765\\ 43210987654321098765432109876543210987654321098765432\\ 10987654321987654321098765432109876543210987654321098\\ 76543210987654321098765432109876543210987654321098765\\ 43210876543210987654321098765432109876543210987654321\\ 09876543210987654321098765432109876543210987654321097\\ 65432109876543210987654321098765432109876543210987654\\ 3210987654321098765432109876543210987654321098\end{tabular} \\ \hline
	\end{tabular}
\end{table}

\subsection{Configuration, Reachability and Reversibility}
\begin{figure}[h]
	\centering
\vspace{-1em}
	\includegraphics[scale=0.3]{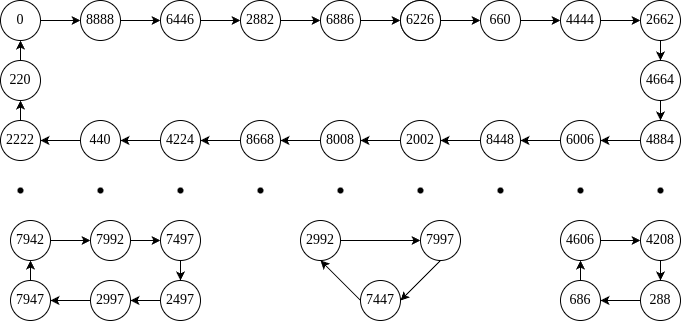}
	\caption{Configuration transition diagram for cell length= $4$, state=$10$ reversible first degree CA $\langle 0,0,0,0,1,0,1,8 \rangle$} 
	\label{fig1}
\vspace{-1em}
\end{figure}

The present state of all cells at a given time is called the configuration of the CA. The evolution of a CA is determined by a global transition function $G_n$ such that $G_n: C_n \rightarrow C_n$ where $C_n =$S$^n$ represents the configuration space of an $n$- cell CA. Let $x, y \in C_n$ be two configurations of the cellular automaton $G_n$. Configuration $y$ is \emph{reachable} from configuration $x$ if there exists an $l_1 \in \mathbb{N}$ such that $G_n^{l_1}(x) = y$; otherwise, $y$ is not reachable from $x$.
Similarly, if $x$ is also reachable from configuration $y$, then they are reachable from each other and they are in the same cycle. These configurations $x$ and $y$ are also called \emph{cyclic} configurations. If there exists no $x \in {C_n}$ such that $G_n^k(x) = y$  for any $K \in \mathbb{N}$, then $y$ is called a \emph{non-reachable} configuration.
Let $C_i \subseteq C_n$ be a set of configurations such that $G_n^l(x) = x$, $\forall x \in C_i$, where $l \in \mathbb{N}$ and $|\mathcal{C}_i| = l$. Then, any $x \in C_i$ is cyclic and reachable from all configurations of $C_i$.
A CA is called reversible, if all configurations are part of some cycles~\cite{mukherjee2021clustering}. All cycles of a reversible CA forms a \emph{cyclic space}. For instance, Figure \ref{fig1} illustrates a snapshot of the evolution of the decimal FDCA $\langle 0, 0, 0, 0, 1, 0, 1, 8 \rangle$ with cell length $4$. This CA has $220$ number of cycles with an average cycle length of $45$, and maximum cycle length of $60$.
As the whole configuration space of this FDCA encompasses only cycles, with no non-reachable configurations, it is a reversible CA.

This reachability of configurations is the most important factor for creating clusters with reversible CAs. Configurations are within the same cluster if they are reachable from each other; otherwise, they belong to different clusters. 
However, as  the number of cycles in the chosen CA directly relates to the number of clusters generated by that CA, it might happen that this number is larger then the desired number of clusters given by the user. To tackle this, the required number of clusters is produced using an iterative three-stage clustering method. At each stage, the clusters from previous stage are merged. 
The merging of these clusters is determined by a \emph{closeness metric}. The process ends when the desired number of clusters is achieved. However, the effectiveness of the process depends on the chosen CA and the closeness metric for merging.

\subsection{Clustering and Flow of Information}
\label{sec:PreliminariesInfoFlow}
In case of clustering, the objects in same cluster are similar to each other. So, in terms of CA, as configuration represent these objects, the configurations belonging to same cycle have to be similar to each other. To assure this, we need CA rules with fewer state changes during evolution between two consecutive configurations of the CA. This occurs when information flows minimally from one configuration to the next, resulting in nearly identical configurations. 
When self replication is present, it implies that the local interactions and update rules in the CA give rise to repeated or recurring patterns that can be observed at different evolution. In a CA with high self-similarity, for example, a specific pattern or configuration of cell states may repeat itself within the context of the automaton's evolution. This repetition can happen at different scales, meaning the same can be seen in a local neighborhood of cells and larger sections of the CA. As a reult, only objects that are comparable are combined in the same cycle. Therefore, our goal is to choose CAs with a low rate of information flow and a high rate of self-replication.

\subsubsection{$P$ Parameter}
\label{sec:Preliminaries: p parameter}
To measure the flow of information, we take help of an existing chaotic parameter, known as $P$ Parameter~\cite{Supreeti_2018_chaos}. This parameter is calculated in two parts - \emph{information propagation} and \emph{information cooking} for both left and right directions.  For our decimal CA the information propagation to the left neighbor of cell $i$ due to update in cell $i$ ($\Lambda_p$) is calculated as -- 
\begin{equation*}
	\Lambda_p  = \frac{1}{10^{2}}\sum_{i=0}^{10^{2}-1}\lambda^{+1}_i
\end{equation*}
where 
\begin{equation*}
	\lambda^{+1}_i = \frac{1}{10(10-1)}\sum_{r,s \in Sibl_i, r \ne s}\delta^{+1}_i(r,s)
\end{equation*}
and 

\begin{equation*}
	\delta^{+1}_i(r,s)  =
	\begin{cases}
		1 & \text{if $R[r] \ne R[s], r,s \in Sibl_i$} \\
		0 & \text{otherwise}
	\end{cases}
\end{equation*}
Similarly, information propagation to the right neighbors of cell $i$ for change in cell $i$ ($\eta_p$) is calculated as -- 
\begin{equation*}
	\eta_p  = \frac{1}{10^{2}}\sum_{i=0}^{10^{2}-1}\lambda^{-1}_i
\end{equation*}
where 
\begin{equation*}
	\lambda^{-1}_i = \frac{1}{10(10-1)}\sum_{r,s \in Equi_i, r \ne s}\delta^{-1}_i(r,s)
\end{equation*}
and 
\begin{equation*}
	\delta^{-1}_i(r,s)  =
	\begin{cases}
		1 & \text{if $R[r] \ne R[s], r,s \in Equi_i$} \\
		0 & \text{otherwise}
	\end{cases}
\end{equation*}
Information cooking from left and right neighbor is computed using the RMTs of $L$-set and $R$-set in order to ascertain the impact of the neighbor state update on cell $i$. Information cooking from the left represents the overall likelihood that cell $i$ will undergo a change in response to the change in its neighboring cell on the left. It is defined as follows:
\begin{equation*}
	\eta_c  = \frac{1}{10^{3}}\sum_{i=0}^{10^{3}-1}\lambda^{+2}_i
\end{equation*}
where 
\begin{equation*}
	\lambda^{+2}_i = \frac{1}{(10-1)^2+1}\sum_{j \in R\text{-set}(i)}\delta^{+2}_i(j)
\end{equation*}
and 
\begin{equation*}
	\delta^{+2}_i(j)  =
	\begin{cases}
		1 & \text{if $R[i] \ne R[j], j \in R\text{-set}(i)$} \\
		0 & \text{otherwise}
	\end{cases}
\end{equation*}
In the same way, information cooking from right neighbor is defined as --
\begin{equation*}
	\Lambda_c  = \frac{1}{10^{3}}\sum_{i=0}^{10^{3}-1}\lambda^{-2}_i
\end{equation*}
where 
\begin{equation*}
	\lambda^{-2}_i = \frac{1}{(10-1)^2+1}\sum_{j \in L\text{-set}(i)}\delta^{-2}_i(j)
\end{equation*}
and 
\begin{equation*}
	\delta^{-2}_i(j)  =
	\begin{cases}
		1 & \text{if $R[i] \ne R[j], j \in L\text{-set}(i)$} \\
		0 & \text{otherwise}
	\end{cases}
\end{equation*}

The overall possibility of creating disturbance to the left neighbor because of change in the current cell is $\mathbf{L}=(\Lambda_p, max(\eta_p, \eta_c))$ and to the right neighbor is $\mathbf{R}=(\eta_p, max(\Lambda_p,\Lambda_c))$. The parameter $P$ for the CA is defined as $$P=max(\mathbf{L},\mathbf{R})$$ where $\mathbf{L}\ge \mathbf{R}$ if $\Lambda_p \langle \eta_p$. 
If $\Lambda_p = \eta_p$, then $\mathbf{L}\ge \mathbf{R}$ if $max(\eta_p, \eta_c) \ge max(\Lambda_p,\Lambda_c)$. In the paper~\cite{Supreeti_2018_chaos}, it has been argued that, a CA can be considered chaotic if $P=(l,r)$ where $l\ge 0.75$ and $r\ge 0.5$. As we need non-chaotic CAs , we have set our filtering criteria based on $\eta_p$, $\eta_c$ , $\Lambda_p$ and $\Lambda_c$


\subsubsection{Self-Information Propagation}
To measure the impact of self-replication on the evolution of the CA, we calculate the \emph{self-information propagation} with the help of the \emph{self-equivalent} RMTs defined as follows:
	\begin{definition}\label{def:equivalent}
	A set of RMTs $r_0, r_1, \cdots, r_{d-1}$ of a $d$-state $3$-neighborhood CA are called \emph{self-equivalent}, for any $r_i,r_j$, where $r_i \ne r_j$, $r_i \equiv r_j \pmod{d}$.
	\end{definition} 
	For all $i$, $0\le i\le d^2-1$, the set of self-equivalent RMTs $Self_i=\{s~|~ s= d\times k+i, 0\le k \le d-1\}$.  That is, in case of our decimal CAs, $Self_0=\{0,10,20, \cdots, 90\}$, $Self_1=\{1,11,21, \cdots, 91\}$, and so on up to $Self_{99}$.
	
		To obtain the self information propagation, we define a binary function $\delta^0_i$ where
	\begin{equation*}
		\delta^0_i(r,s)  =
		\begin{cases}
			1 & \text{if ${R}[r] \ne {R}[s]$, $r,s \in Self_i$} \\
			0 & \text{otherwise}
		\end{cases}
	\end{equation*}
 Consequently, the overall possibility of self-information propagation due to change in the current cell is
		\begin{align*}
		\Delta_p  = \frac{1}{d^{2}}\sum_{i=0}^{d^{2}-1}\lambda^0_i
	\end{align*}
	where
	\begin{align*}
		\lambda^0_i = \frac{1}{d(d-1)}\sum_{r,s \in Sibl_i, r \ne s}\delta^k_i(r,s)
	\end{align*}
	Here, $d=10$ and $0\le i \le d^2-1=99$. If RMTs of $Self_i$ have different next state values for all $i$, then $\Delta_p$ approaches to the maximum value $1$. Here, we consider the CAs with atleast 50\% self-information propagation rate.

\subsection{Clustering Metrics and Benchmarks}
\label{Clustering Metrices and Benchmarks}
Clustering metrics and benchmarks play a crucial role in assessing the quality and performance of the clustering algorithms. These metrics offer quantitative measures of the alignment between the clusters generated by an algorithm and the expected or ground truth clustering structure. There are several commonly employed clustering metrics and benchmarks. It is vital to select appropriate clustering metrics and criteria based on the characteristics of the data and the specific goals of the clustering task. Utilizing a variety of metrics is advisable to gain a comprehensive understanding of clustering performance, as different metrics may be more suitable for different scenarios.
We compare the outcomes achieved with our technique to those obtained using the data mining clustering algorithms such as, K-Means~\cite{hartigan1979algorithm} and Hierarchical~\cite{Hierarchical}, DBSCAN~\cite{ester1996density}, PAM~\cite{JSSv025i04}, Meanshift~\cite{comaniciu2002mean}, Birch~\cite{zhang1997birch} and the existing binary CA based clustering algorithm~\cite{abhishek2023cellular}. The comparison is based on evaluation metrics including the DB score~\cite{ester1996density}, Silhouette score~\cite{silhouette}, and Calinski-Harabasz (CH) score~\cite{calinski1974dendrite} and Dunn Index~\cite{dunn}.

The Silhouette score is a metric used to determine the quality of clustering in data analysis. It measures how similar an object is to its own cluster (cohesion) compared to other clusters (separation). The score ranges from -1 to 1, where a high value indicates that the object is well matched to its own cluster and poorly matched to neighboring clusters. A score close to 0 suggests overlapping clusters. It is commonly used to assess the appropriateness of clustering algorithms and to determine the optimal number of clusters in a dataset. The silhouette score is calculated as follows:
\begin{equation}
	\frac{b-a}{max(a,b)}
\end{equation}
The Davis Bouldin (DB) score is computed by dividing the intra-cluster distances by the inter-cluster lengths. Additionally, similarity is calculated by dividing the intra-cluster distances by the inter-cluster lengths. This calculation can be expressed using the following formula:
\begin{equation}
	\frac{1}{k} \sum _{i=1}^{k} \max _{i \neq j} \frac{\Delta(X_i) + \Delta(X_j)}{\delta(X_i, X_j)}
\end{equation}
where $\delta(X_i, X_j)$ is the inter-cluster distance, representing the distance between cluster $X_i$ and $X_j$, and $\Delta(X_i)$ is the intra-cluster distance of cluster $X_i$, indicating the distance within the cluster $X_i$.
The Calinski-Harabasz (CH) score is a measure of the within-cluster dispersion compared to between-cluster dispersion. It is calculated as:

\begin{equation}
	\frac{\frac{\sum_{k=1}^{K} n_{k}\left\|c_{k}-c\right\|^{2}}{K-1}}{\frac{\sum_{k=1}^{K} \sum_{i=1}^{n_{k}}\left\|d_{i}-c_{k}\right\|^{2}}{N-K}}
\end{equation}
where $d_{i}$ represents one of the features of the dataset $D$, $n_{k}$ and $c_{k}$ are the number of points and centroid of cluster $k$ respectively, $c$ is the global centroid, and $N$ is the total number of data points. A higher value of the CH index indicates that the clusters are dense and well-separated.

The Dunn index is utilized to identify clusters that are compact, exhibiting minimal variance among their members, and well-separated, with a substantial distance between the means of different clusters relative to the variance within each cluster. This index is defined as:

\begin{equation}
	\frac{\min _{1 \leq i \leq j \leq m} \delta\left(C_{i}, C_{j}\right)}{\max _{1 \leq k \leq m} \Delta_{k}}
\end{equation}
Where $\delta\left(C_{i}, C_{j}\right)$ represents the inter-cluster distance, $\Delta_{k}$ denotes the intra-cluster distance, and $m$ signifies the number of clusters.

\subsubsection{Datasets}
\label{Datasets: Clustering Metrices and Benchmarks}
Various datasets with quantitative attributes have been utilized in our paper to analyze the results of clustering algorithms. These datasets (DS) are sourced from \url{http://archive.ics.uci.edu/ml/index.php}.
The details of each dataset are presented in Table~\ref{datasetTable}. 

\begin{table}[!h]
\vspace{-1em}
	\caption{Dataset Description}\label{datasetTable}
	\centering
	\small
	\setlength\tabcolsep{4pt}
		\begin{tabular}{|l|l|l|l|}
			\hline
			No. of Dataset &	Name of Dataset  & \#Instances & \#Features\\
			\hline
			DS1 &	Seeds &	199 & 8 \\ 
			\hline  
			DS2 &	User Knowledge Modelling &	258 & 6\\
			\hline
			DS3 & Heart Failure Clinical Record &	299 & 13 \\
			\hline
			DS4 &	BuddyMove &	249 & 7\\ 
			\hline
	\end{tabular}
\end{table}

\section{The Encoding Technique}
\label{sec:Encoding}
In the context of clustering, encoding techniques refer to the methods used to transform data points or features into a format suitable for clustering algorithms. These techniques involve converting the original data into a numerical or symbolic representation that effectively captures the necessary information for clustering analysis. The selection of an encoding technique depends on various factors such as data characteristics, the clustering algorithm utilized, and the desired attributes of representation. It is crucial to assess data requirements and choose an encoding method that preserves relevant information for effective clustering. 

When dealing with high-dimensional datasets, the need to compress dataset features into an encoded string arises. Simply concatenating feature values into a string may result in an excessively large string length, leading to inefficient computational processes. Therefore, an encoding function is needed to encode the features.
One approach is to use an encoding technique based on the range of feature values to map data objects to encoded strings. While this technique effectively reduces the length of the string, it may also result in data loss and potentially lead to poor clustering outcomes. 
For example, if we have a hypothetical dataset about a set of fruits where each fruit has properties such as the number of nutritional contents, nutritional values, and texture, the first two features have numerical values in them, while the third feature is categorical, either hard or soft. Table~\ref{tab4.1} is the dataset, consisting of 10 instances. If we directly append the decimal numbers of each column to get a decimal string per row, the length of the string may become very large making the scheme inconvenient for high-dimensional dataset. Similarly, if we use range based encoding technique, then the numerical attributes are encoded in ranges such as Number of Nutritional contents is divided into {[}4, 11{]},{[}12, 30{]} and {[}31,40{]} represented by 00, 01 and 11 respectively. Similarly, the other attribute is encoded. Categorical data are encoded as 01 for Hard and 10 for Soft. For a smaller number of features, this encoding technique will work but when we have a greater number of features which is the case for Big Data then the encoded configuration length can be very large. Furthermore, if unrelated data points are encoded to the same string, it can adversely affect clustering results, as these points may belong to different clusters. To address this issue, it is important to utilize a unique encoding method to ensure that two data points share the same encoding only if they are intended to be placed in the same cluster.

\begin{sidewaystable}[htbp]
	\centering
\vspace{-1em}
	\caption{Range based encoding of the dataset features}\label{tab4.1}
		\begin{tabular}{c|c|c|c|c|c|c|c}
			\hline
			Data Object & \multicolumn{4}{c|}{Numerical Attributes} & \multicolumn{2}{c|}{Categorical Attribute} & Encoded Configuration\\& Nutritional Contents & Encoding & Nutritional value & Encoding & Texture & Encoding \\
			\hline
			1 & 30 & 01 & 9 & 11 & Hard & 01 & 011101 \\
			2 & 32 & 11 & 8 & 11 & Soft & 10 & 111110 \\
			3 & 4 & 00 & 9.5 & 11 & Soft & 10 & 001110 \\
			4 & 20 & 01 & 4 & 00 & Hard & 01 & 010001 \\
			5 & 12 & 01 & 4.5 & 01 & Soft & 10 & 010110 \\
			6 & 6 & 00 & 7 & 01 & Hard & 01 & 000101 \\
			7 & 31 & 11 & 6.8 & 01 & Soft & 10 & 110110 \\
			8 & 11 & 00 & 3 & 00 & Soft & 10 & 000010 \\
			9 & 40 & 11 & 2.6 & 00 & Soft & 10 & 110010 \\
			10 & 35 & 11 & 9.3 & 11 & Soft & 10 & 111110 \\
			\hline
	\end{tabular}
\vspace{-1em}
	
\end{sidewaystable} 

For this reason, in Ref.~\cite{narodia2023godel}, an attempt was made to use non-cryptographic hash-based encoding, which resulted in poor clustering scores. Subsequently, a new encoding technique based on G{\"o}del numbering was introduced which we will utilize here. Here, for the sake of completeness, we discuss the G{\"o}del number-based encoding scheme and the benefits of using this scheme for our clustering algorithm.

\subsection{G{\"o}del Number Based Encoding}
\label{sec:GodelEncoding}
G{\"o}del numbering is a classical encoding method of representing symbols and well-formed formula (wffs) of a formal language by a unique natural number, called the \emph{G{\"o}del Number}. Introduced by Kurt G{\"o}del in 1931, to prove the incompleteness theorems, this works on the essence of the unique prime factorization theorem, the fundamental theorem of arithmetic. The G{\"o}del number of a sequence of positive integers $(g_1, g_2, \ldots, g_n)$ is obtained by computing the product of the first $n$ prime numbers raised to the power of their respective values in the sequence. According to the unique prime factorization theorem, any number can be uniquely expressed as a product of prime numbers. For example, any number $\mathcal{G}$ can be uniquely represented as the product of prime numbers raised to specific powers, as shown below:
\begin{equation}
	\label{Godel Equation}
	\mathcal{G} = 2^{g_1} * 3^{g_2} * 5^{g_3} * \ldots * a_{n}^{g_n}
\end{equation}
where $2,3,5, \ldots, a_n$ are the sequence of the first $n$ prime numbers and $g_1, g_2, \ldots, g_n$ are the natural numbers. As an instances, the number $900$ is encoded as $2^2 * 3^2 * 5^2$. Therefore, given the sequence $(g_1, g_2, \ldots, g_n)$, the G{\"o}del number associated with this sequence is $\mathcal{G}$ which can be decoded back to this $(g_1, g_2, \ldots, g_n)$. That means, G{\"o}del numbering is completely reversible, that is, decoding the original sequence from the G{\"o}del number is possible.

In this work, by G{\"o}del numbering based encoding or G{\"o}del encoding, we refer to the technique that applies G{\"o}del numbering to represent and analyze data in the context of clustering algorithms. In G{\"o}del number based clustering, data points or objects are represented using G{\"o}del numbers, which are unique numerical representations obtained through the systematic encoding process of G{\"o}del numbering. These numbers represent the attributes, features, and structures of the data points, effectively encapsulating their relevant information. Subsequently, the clustering process operates on these G{\"o}del numbers rather than the original dataset. 

Although classically G{\"o}del numbering is applicable for symbols and wffs of formal languages, to apply this encoding for clustering, we consider the sequence of values of all features of a data object as the sequence for G{\"o}del numbering~\cite{narodia2023godel}. However, these feature values can be a real number whereas G{\"o}del numbering works for only positive integers. So to apply G{\"o}del numbering over a real-life dataset, first we need to remove the decimal point from the feature values by applying scale up process. This obviously do not violate the feature properties. 

Let us consider a hypothetical data set with four features to compute the G{\"o}del numbers based encoding as shown in Table~\ref{tab:my-table}.
\begin{table}[hbtp]
	\centering
	\caption{Hypothetical Data set}
	\label{tab:my-table}
	\begin{tabular}{|l|l|l|l|l|l|}
		\hline
		\textbf{Feature} & \textbf{F1} & \textbf{F2} & \textbf{F3} & \textbf{F4} & \textbf{Gödel Number} \\ \hline
		\textbf{a} & 10 & 2 & 1 & 7 & 37948861440    \\ \hline
		\textbf{b} & 10 & 3 & 5 & 6 & 10164873600000 \\ \hline
		\textbf{c} & 5  & 5 & 4 & 2 & 238140000      \\ \hline
		\textbf{d} & 2  & 1 & 5 & 9 & 1513260262500  \\ \hline
	\end{tabular}
\end{table}
We now apply G{\"o}del encoding technique over this data set. For this, we need the first four primes 2,3,5, and 7. Then, the object $a$ can be represented as: 
$\bar{a} = 2^{10} * 3^{2} * 5^{1} * 7^{7}$. Similarly, the other objects can be encoded as follows: \\
$\bar{b} = 2^{10} * 3^{3} * 5^{5} * 7^{6}$,\\
$\bar{c} = 2^{5} * 3^{5} * 5^{4} * 7^{2}$, and \\
$\bar{d} = 2^{2} * 3^{1} * 5^{5} * 7^{9}$. \\
Over this encoded dataset, decimal FDCA based clustering can be applied.

\subsection{Benefits of the Scheme}
\label{sec:SortGodel}
The G{\"o}del number-based encoding ensures that each formal expression can be computed algorithmically and has a unique numerical representation. We aim to verify whether our G{\"o}del number-based encoding preserves the characteristics of the features of the given dataset for clustering. To achieve this, we undertake the following actions:
\begin{enumerate}
	\item \textbf{Generation:} For every row in the dataset, generate a G{\"o}del number.
	\item \textbf{Sorting:} Depending on their value, arrange the G{\"o}del numbers in ascending order.
	\item \textbf{Finding Gaps:} Given $k$, the desired number of clusters, find the $k-1$ greatest gaps in the distribution of these numbers.
	\item \textbf{Clustering:} The rows that correspond to the G{\"o}del numbers in each gap are part of the same cluster.
\end{enumerate}

We can assess the effectiveness of G{\"o}del encoding compared to other algorithms by sorting the numbers in ascending order based on their values and attempting to divide them into the number of clusters after generating G{\"o}del numbers for each row. 
Identifying the first $k-1$ greatest distances between consecutive elements enables us to segregate the data into the desired number of clusters, denoted as $k$. In this paper, we term this technique as \emph{Sort Gödel}. Subsequently, these clusters undergo normal evaluation metric testing.


\begin{sidewaystable}[htbp]
	\centering
	\vspace{-1em}
	\caption{Comparison of hashing techniques based on merging metrics}
	\label{Comparison of Tech}
		\begin{tabular}{|c|c|c|c|c|c|}
			\hline
			\multicolumn{2}{|c|}{\multirow{2}{*}{Dataset}} & \multicolumn{4}{c|}{Evaluation Metrics} \\
			\hhline{~~----}
			\multicolumn{2}{|c|}{} & Silhouette Score & Davis Bouldin & Calinski Harabasz & Dunn Index \\
			\hline
			\multirow{2}{*}{Seeds} & K-Means & 0.531 & 0.659 & 345.488 & 0.0095 \\
			\hhline{~-----}
			& Sort Gödel & 0.493 & 0.824 & 318.944 & 0.0072 \\
			\hline
			\multirow{2}{*}{User Knowledge Modelling} & K-Means & 0.265 & 0.659 & 345.488 & 0.1279 \\
			\hhline{~-----}
			& Sort Gödel & 0.203 & 0.993 & 22910.758 & 0.0022 \\
			\hline
			\multirow{2}{*}{Clinical Heart Failure} & K-Means & 0.583 & 0.659 & 266.924 & 0.1285 \\
			\hhline{~-----}
			& Sort Gödel & 0.383 & 0.777 & 58.740 & 0.0035 \\
			\hline
			\multirow{2}{*}{BuddyMove} & K-Means & 0.320 & 1.336 & 119.97 & 0.045 \\
			\hhline{~-----}
			& Sort Gödel & 0.233 & 0.962 & 51.456 & 0.0080 \\
			\hline
		\end{tabular}
	\vspace{-1em}
\end{sidewaystable}

The comparison results between this approach and the K-Means algorithm are presented in Table~\ref{Comparison of Tech}. For example, when constructing the Gödel number for the \textit{seeds} dataset using the above approach, the silhouette score for that dataset is 0.4934, which is remarkably close to the K-Means score of 0.531. This suggests that the attributes of the features can be effectively preserved using the Gödel number-based encoding function.

\subsection{Restrictions and how to resolve}
The G{\"o}del number can grow significantly in length based on the number of features, necessitating the division of the extensive G{\"o}del number into smaller segments to tackle this challenge. This entails transforming the original dataset into a new data frame upon which clustering techniques are subsequently applied. Various methods can be taken for constructing a G{\"o}del data frame, as outlined in Ref.~\cite{narodia2023godel}. The G{\"o}del numbers generated by each scheme exhibit variability for the same object and can vary in length. Here, we consider the following scheme:
\begin{enumerate}
	\item Pre-process to remove decimal points from the dataset.
	\item Generate the Gödel number for all data objects, considering all features of the dataset.
	\item Consider these numbers as a decimal strings.
	\item If length of the strings for different objects are varied, append zeros in the beginning of such strings to make them of equal length.
	\item Then split the string vertically into a number of substrings, where this split size is as per the user's choice or the computational limit of the machine.
\end{enumerate}

\begin{figure}[!h]	
	\vspace{-1em}
	\centering
	\subfloat[Step 2 \label{godel1}]{%
		\includegraphics[width=3.5cm,height=4.5cm]{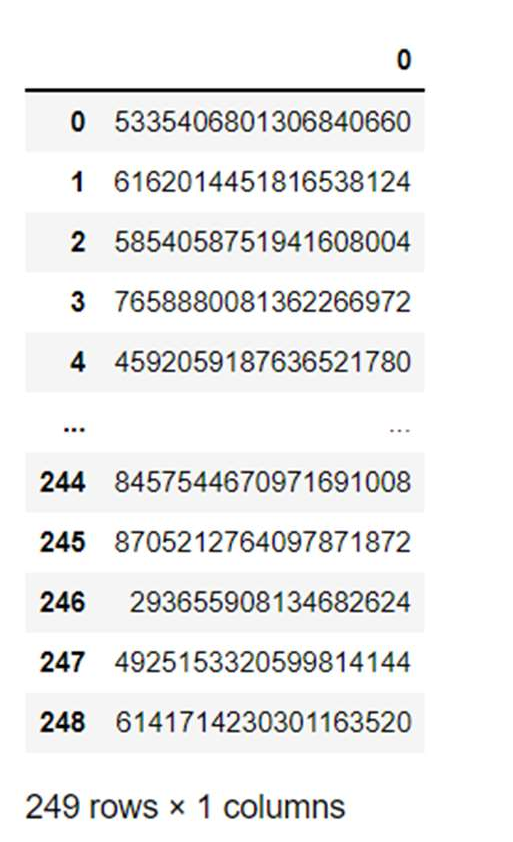}}\hfill
	\subfloat[Step 5\label{godel2}]{%
		\includegraphics[width=3.5cm,height=4.5cm]{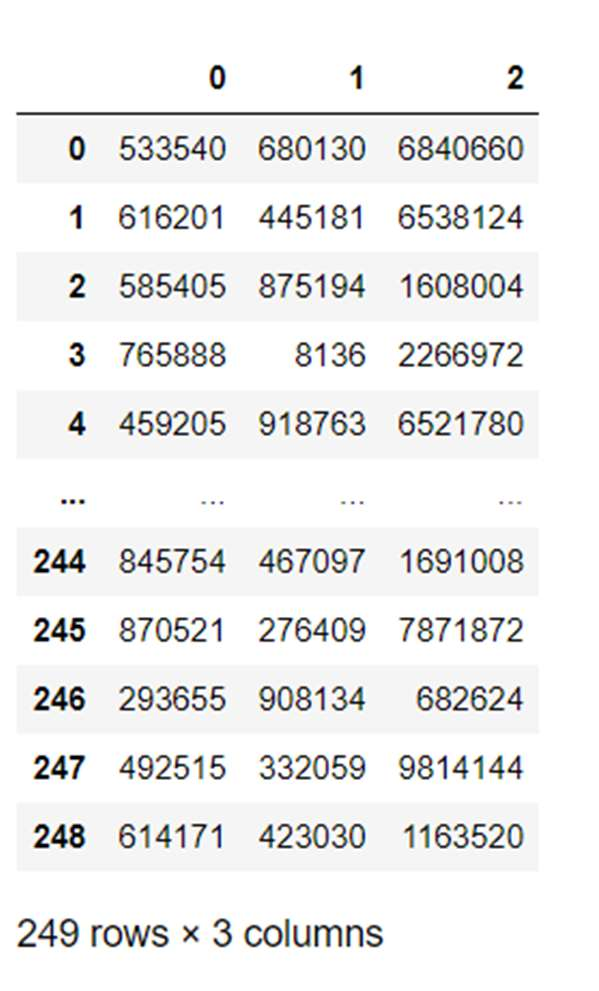}}\hfill
	\caption{G{\"o}del number based encoding scheme over\emph{BuddyMove} dataset}	\label{fig:godel}
	\vspace{-1em}
\end{figure}

Figure~\ref{fig:godel} illustrates this approach over the \emph{buddymove} dataset. Here, (Figure~\ref{godel2}), the original G{\"o}del number generated per object is divided into three parts to generate three substrings (Figure~\ref{godel2}). Now, for comparison with the state-of-the-art algorithms (after G{\"o}del encoding), we may consider each split as a distinct feature.

The encoding function based on G{\"o}del numbers assists in maintaining the characteristic of the feature value. By encoding using G{\"o}del numbering scheme, these numerical feature values are combined and converted into a decimal number. As G{\"o}del numbering based encoding generates decimal number, this completely fits our scheme. To cluster the encoded dataset using G{\"o}del number based encoding, opting for reversible decimal cellular automata is a natural choice over binary CAs. This decision aids in preventing data loss and preserves the features of the encoding. 
In the next section, we select the decimal CAs which suits our requirement for clustering based on some theoretical properties.

\section{Selection of Proper Rules}
The preceding section demonstrates how to utilize G{\"o}del numbering to encode any dataset such that reversible first degree cellular automata can be applied for clustering. G{\"o}del number based encoding combines numerical feature values and converts them into a decimal number, ensuring no data loss and maintaining the properties of the features. In clustering problems, features are typically interdependent and considered together when creating clusters. Therefore, when using FDCA based clustering, it is important to find the potential candidate FDCA rules which do not depend solely on a single feature. However, exhaustively searching the whole $10^8$ rules to find the best candidate is computationally very costly. So specific criteria are needed to identify the suitable candidate FDCA rules for clustering.

\subsection{Selection of 1560 initial rules}
\begin{figure}[h]
	\centering
	\vspace{-1em}
	\includegraphics[scale=0.45]{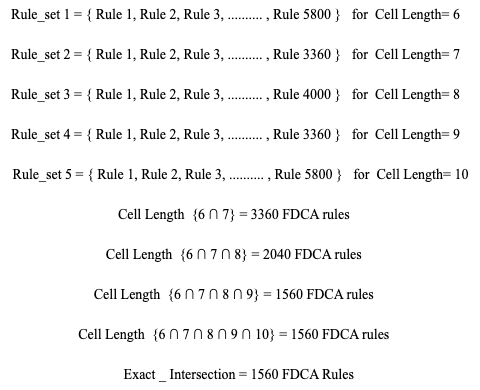}
	\caption{Exact intersection 1560 reversible FDCA rules} \label{Exactintersection}
	\vspace{-1em}
\end{figure}

Our goal is to find FDCA rules that minimize the intra-cluster distance among configurations in the same cycle. For that, we need to identify rules that demonstrate a strict locality property. To discover reversible rules, a thorough analysis of all potential rules is required. In this work, we are employing a split size of $6$ to $10$ for our scheme. So, we consider the cell length $n \in \{6, 7, 8, 9, 10\}$. We specifically select only those decimal FDCAs under null boundary condition which are reversible for each $n \in \{6, \cdots, 10\}$. The possible number of decimal reversible CAs are presented in Table~\ref{Selected 1560 Rules} for the respective cell lengths. Out of the set of rules under consideration, a subset comprising 1560 rules which are common for every cell length $n \in \{6, \ldots, 10\}$ are selected for subsequent processing. The process of selecting these rules from set of possible rules (Table \ref{Selected 1560 Rules}) is shown in Figure~\ref{Exactintersection}.

\begin{table}[hbtp]
	\centering
	\caption{Selected rules from 6 to 10 cell length for 10-state FDCA}
	\label{Selected 1560 Rules}
	\small
	\begin{tabular}{|r|r|}
		\hline
		\multicolumn{1}{|l|}{Cell length } & \multicolumn{1}{l|}{Generated FDCA rules for state 10} \\ \hline
		6                            & 5800                                   \\ \hline
		7                            & 3360                                   \\ \hline
		8                            & 4000                                   \\ \hline
		9                            & 3360                                   \\ \hline
		10                           & 5800                                   \\ \hline
	\end{tabular}
	\vspace{-1em}
\end{table}

\subsection{Rule Selection Criterion}
\label{Rule Selection Criterion}
The process of selecting rules in FDCA for clustering is often iterative and exploratory, driven by the desired clustering objectives and insights gained from observing the emergent behavior of the automaton. In order to narrow down the search space and identify rules that align optimally with our criteria, additional theoretical filtering criteria are applied to the selected 1560 rules. This further refinement is based on assessing the chaotic parameters and analyzing the cyclic spaces associated with each rule for each cell length. In pursuit of our current objective, the selection of decimal FDCA rules is directed towards minimizing changes during evolution. Here, we present two criteria for selecting rules:
\begin{enumerate}[label=(\alph*),noitemsep]
	\item Based on the cyclic structure of the rules.
	\item Based on chaotic parameters.
\end{enumerate}

\subsection{Filtering Criteria I: Based on the cyclic structure of the rules}
\label{criteria I: Cyclic structure of the rules}
As previously explained, our clustering approach using decimal first degree cellular automata (FDCAs) relies on identifying clusters within the cyclic space of the cellular automaton dynamics. This methodology, termed cycle-based clustering, ensures the preservation of cycles with smaller intra-cluster distances. While cellular automata have demonstrated effectiveness in clustering tasks based on their cyclic space, identifying an FDCA capable of adequately partitioning objects into the desired number of clusters can be challenging. Moreover, our method for selecting significant FDCA rules is compatible with a reduced rule set, making the number of cycles an important parameter in clustering.
Our first criteria for selecting rules is based on the number of cycles. Out of the $1560$ rules initially selected, we extract 480 rules with the minimum number of cycles for the cell length, specifically $72$, $76$, and $128$ cycles, respectively. To further refine our selection, we analyze these $480$ rules across cell lengths ranging from $6$ to $9$. By considering parameters such as the number of cycles and the maximum cycle length, we narrow down the set to $28$ rules from the initial $480$. These $28$ selected rules are chosen based on their favorable performance according to the specified criteria.
For example, in the context of rule $\langle 0, 0, 0, 0, 1, 7, 8, 1 \rangle$ with cell length of $6$, it is observed that this rule create $72$ cycles, with a maximum cycle length of $15624$. This observation remains consistent when compared to all other rules listed in Table \ref{Char 28 rule} for their respective cell length as $6$. This Table~\ref{Char 28 rule} presents the characteristics of the selected $28$ rules, categorized by different cell lengths spanning from \( n=6 \) to \( n=9 \). These characteristics encompass both the number of cycles and the maximum cycle length associated with each distinct cell length. Across the cell lengths of $6$ through $9$, an intriguing consistency emerges: all 28 selected rules consistently exhibit same number of cycles, irrespective of the cell length. However, the maximum cycle length varies across the different cell lengths. This observation underscores the unique behavior of the selected rules under varying cell lengths, suggesting potential insights into their clustering behavior within the context of the study.

\begin{table}[hbtp]
	\centering
	\vspace{-1em}
	\caption{28 Selected Rules based on cyclic structure analysis}
	\label{28rules}
	\begin{tabular}{|cccc|}
		\hline
		\multicolumn{4}{|c|}{\textbf{28 Rules Selected based on cyclic structure}}                                                                     \\ \hline
		\multicolumn{1}{|c|}{$\langle 0,0,0,0,1,7,8,1\rangle$} & \multicolumn{1}{c|}{$\langle0,0,0,0,3,7,6,9\rangle$} & \multicolumn{1}{c|}{$\langle0,0,0,0,4,9,3,7\rangle$} & $\langle0,0,0,0,3,7,6,3\rangle$ \\ \hline
		\multicolumn{1}{|c|}{$\langle0,0,0,0,1,7,8,3\rangle$} & \multicolumn{1}{c|}{$\langle0,0,0,0,3,9,4,1\rangle$} & \multicolumn{1}{c|}{$\langle0,0,0,0,4,9,3,9\rangle$} & $\langle0,0,0,0,3,7,6,7\rangle$ \\ \hline
		\multicolumn{1}{|c|}{$\langle0,0,0,0,1,7,8,7\rangle$} & \multicolumn{1}{c|}{$\langle0,0,0,0,3,9,4,3\rangle$} & \multicolumn{1}{c|}{$\langle0,0,0,0,6,7,3,1\rangle$} & $\langle0,0,0,0,3,7,6,9\rangle$ \\ \hline
		\multicolumn{1}{|c|}{$\langle0,0,0,0,1,7,8,9\rangle$} & \multicolumn{1}{c|}{$\langle0,0,0,0,3,9,4,7\rangle$} & \multicolumn{1}{c|}{$\langle0,0,0,0,6,7,3,3\rangle$} & $\langle0,0,0,0,3,9,4,1\rangle$ \\ \hline
		\multicolumn{1}{|c|}{$\langle0,0,0,0,3,7,6,1\rangle$} & \multicolumn{1}{c|}{$\langle0,0,0,0,3,9,4,9\rangle$} & \multicolumn{1}{c|}{$\langle0,0,0,0,6,7,3,7\rangle$} & $\langle0,0,0,0,3,9,4,3\rangle$ \\ \hline
		\multicolumn{1}{|c|}{$\langle0,0,0,0,3,7,6,3\rangle$} & \multicolumn{1}{c|}{$\langle0,0,0,0,4,9,3,1\rangle$} & \multicolumn{1}{c|}{$\langle0,0,0,0,6,7,3,9\rangle$} & $\langle0,0,0,0,3,9,4,7\rangle$ \\ \hline
		\multicolumn{1}{|c|}{$\langle0,0,0,0,3,7,6,7\rangle$} & \multicolumn{1}{c|}{$\langle0,0,0,0,4,9,3,3\rangle$} & \multicolumn{1}{c|}{$\langle0,0,0,0,3,7,6,1\rangle$} & $\langle0,0,0,0,3,9,4,9\rangle$ \\ \hline
	\end{tabular}
		\vspace{-1em}
\end{table}

\begin{table}[hbtp]
	\centering
	\vspace{-1em}
	\caption{Characteristics of Selected 28 Rules}
	\label{Char 28 rule}
	\begin{tabular}{|ccccc|}
		\hline
		\multicolumn{5}{|c|}{\textbf{Cyclic based selected 28 Rules}} \\ \hline
		\multicolumn{1}{|c|}{Cell Length}          & \multicolumn{1}{c|}{n=6}   & \multicolumn{1}{c|}{n=7}  & \multicolumn{1}{c|}{n=8}   & n=9     \\ \hline
		\multicolumn{1}{|c|}{No. of Cycles}        & \multicolumn{1}{c|}{72}    & \multicolumn{1}{c|}{6800} & \multicolumn{1}{c|}{9744}  & 4181152 \\ \hline
		\multicolumn{1}{|c|}{Maximum cycle length} & \multicolumn{1}{c|}{15624} & \multicolumn{1}{c|}{1560} & \multicolumn{1}{c|}{10416} & 240     \\ \hline
	\end{tabular}
	\vspace{-1em}
\end{table}

\subsection{Filtering Criteria II: Chaotic parameter }
\label{criteria II: Chaotic parameter}
The dynamics of cellular automata (CA) systems are intricately governed by the chaotic parameter, whose varying values elicit diverse behaviors, including regular, chaotic, or complex dynamics. These dynamic attributes play a pivotal role in shaping the formation of clusters within the CA. The behavior of individual cells in this context exhibits temporal unpredictability, contributing to the consequential effects on cluster formation and evolution within the CA. The introduction of chaotic behavior into the CA system adds a layer of complexity to the clustering process, resulting in clusters manifesting intricate patterns, irregular shapes, and varying densities.
To measure this, we use $P$ Parameter which is calculated in two parts: information propagation and information cooking (see Section~\ref{sec:Preliminaries: p parameter}).
In the context of clustering, this parameter plays a significant role in shaping how clusters of configurations form. A higher value of $P$ accelerates the diffusion of information across configuration space of a CA. This faster diffusion can potentially disrupt the formation of clusters with similar elements. Conversely, a lower value of $P$ slows down the diffusion process, allowing clusters to form over similar configurations. By systematically adjusting the parameter $P$ and observing its effects on clustering patterns, researchers can gain valuable insights into the underlying dynamics of the system.
Moreover, applying restrictive filtration criteria to reduce the rule space enhances our ability to analyze the impact of different parameter values on clustering. This approach enables a more comprehensive exploration of the parameter space, leading to a deeper understanding of how different parameters influence the formation of clusters within first degree cellular automata.

\begin{table}[hbtp]
	\vspace{-1em}
	\caption{Sample rules from 1560 selected rules}
	\label{IFilter:280 Rules}
		\begin{tabular}{|c|c|c|c|c|c|c|}
			\hline
			\textbf{FDCA Rules} & \textbf
			{\begin{tabular}[c]{@{}c@{}}$\Lambda_p$\end{tabular}} & \textbf
			{\begin{tabular}[c]{@{}c@{}}$\eta_p$\end{tabular}} & \textbf
			{\begin{tabular}[c]{@{}c@{}}$\Lambda_c$\end{tabular}} & \textbf
			{\begin{tabular}[c]{@{}c@{}}$\eta_c$\end{tabular}} & \textbf{P-value}            & \textbf
			{\begin{tabular}[c]{@{}c@{}}$\Delta_p$ \end{tabular}} \\ \hline
			$\langle 0, 0, 0, 0, 0, 1, 0, 0 \rangle$ & 0.0 & 0.0& 0.98779666& 0.98779666& {(}0.0, 0.98779666{)}& 0.6444442\\ \hline
			$\langle 0, 0, 0, 0, 0, 1, 5, 0 \rangle$ & 0.5555562& 0.0& 0.98779666& 0.92682225& {(}0.5555562, 0.98779666{)} & 0.6444442\\ \hline
			$\langle 0, 0, 0, 0, 3, 9, 0, 9 \rangle$ & 0.0 & 1.0 & 0.87805146 & 0.98779666& {(}0.0, 1.0{)} & 0.6444442\\ \hline
			$\langle 0, 0, 0, 0, 1, 9, 0, 0 \rangle$ & 0.0& 1.0& 0.87805146& 0.98779666& {(}0.0, 1.0{)} & 0.591111\\ \hline
			$\langle0, 0, 0, 0, 5, 1, 0, 0 \rangle$ & 0.0& 0.5555562& 0.92682225& 0.98779666& {(}0.5555562, 0.98779666{)} & 0.6444442                                                                        \\ \hline
		\end{tabular}
	\vspace{-1em}
\end{table}
We want to select non-chaotic rules with good self information propagation (or, self similarity) rate such that the generated cycles (clusters) have good chance of having only similar configuration as part of the same cluster. To do this, our filtering process is considers $\Delta_p \ge 0.5$ (Self Information Propagation). Then we further refine the selection of rules based on the chaotic parameters $\Lambda_p$ (Left Information Propagation),  $\eta_p$ (Right Information Propagation), $\Lambda_c$ (Left Information Cooking), $\eta_c$ (Right Information Cooking). Following are the detailed steps:

\begin{itemize}
\item[--] \textbf{First Filter}: In the first filtering criteria, we select 280 rules by setting the value of the first argument (L) of $P$ to be less than or equal to 0.7, and ensuring that the self-information propagation is greater than or equal to 0.5. Out of the initial pool of 1560 rules, only 280 rules meet these criteria based on chaotic parameters. A sample rules meeting the criteria are shown in Table \ref{IFilter:280 Rules} where each row in the table corresponds to a specific rule, indicating the values of information propagation and cooking for left and right neighbors, along with statistical measures such as P-value and self-information propagation.
	
\item[--] \textbf{Second Filter}: In the second filtering criteria, we set the conditions $!(\Lambda_p = 0, \eta_p = 0)$ and $!(\Lambda_p = 0, \eta_p = 1)$ and $!(\Lambda_p = 1, \eta_p = 0)$. This set of filtering criteria filters rules based on three specified conditions:
\begin{enumerate}[label=\alph*),noitemsep]
		\item $!(\Lambda_p = 0, \eta_p = 0)$: Both $\Lambda_p$ and $\eta_p$ cannot be simultaneously equal to 0.
		\item $!(\Lambda_p = 0, \eta_p = 1)$: Both $\Lambda_p$ and $\eta_p$ cannot be simultaneously equal to 0 and 1, respectively.
		\item $!(\Lambda_p = 1, \eta_p = 0)$: Both $\Lambda_p$ and $\eta_p$ cannot be simultaneously equal to 1 and 0, respectively. \newline
	\end{enumerate}
After applying this filter, 80 rules remain out of the $1560$ rules  that satisfy these conditions.
A list of these rules is presented in Table~\ref{IIFilter:80Rules}.

\begin{table}[hbtp]
	\centering
		\vspace{-1em}
	\caption{Second Filter (80 Selected FDCA Rules): $!(\Lambda_p = 0, \eta_p = 0)$, $!(\Lambda_p = 0, \eta_p = 1)$, and $!(\Lambda_p = 1, \eta_p = 0)$}
	\label{IIFilter:80Rules}
		\begin{tabular}{|cccc|}
			\hline
			\multicolumn{4}{|c|}{\textbf{Second Chaotic Filter - 80 Selected FDCA Rule}} \\ \hline
			\multicolumn{1}{|c|}{$\langle0, 0, 0, 0, 0, 1, 5, 0\rangle$} &
			\multicolumn{1}{c|}{$\langle0, 0, 0, 0, 0, 3, 5, 6\rangle$} &
			\multicolumn{1}{c|}{$\langle0, 0, 0, 0, 0, 9, 5, 2\rangle$} &
			{$\langle0, 0, 0, 0, 5, 1, 0, 8\rangle$} \\ \hline
			
			\multicolumn{1}{|c|}{$\langle0, 0, 0, 0, 5, 7, 0, 4\rangle$} &
			\multicolumn{1}{c|}{$\langle0, 0, 0, 0, 0, 1, 5, 1\rangle$} &
			\multicolumn{1}{c|}{$\langle0, 0, 0, 0, 0, 3, 5, 7\rangle$} &
			{$\langle0, 0, 0, 0, 0, 9, 5, 3\rangle$} \\ \hline
			
			\multicolumn{1}{|c|}{$\langle0, 0, 0, 0, 5, 1, 0, 9\rangle$} &
			\multicolumn{1}{c|}{$\langle0, 0, 0, 0, 5, 7, 0, 5\rangle$} &
			\multicolumn{1}{c|}{$\langle0, 0, 0, 0, 0, 1, 5, 2\rangle$} &
		    {$\langle0, 0, 0, 0, 0, 3, 5, 8\rangle$} \\ \hline
			
			\multicolumn{1}{|c|}{$\langle0, 0, 0, 0, 0, 9, 5, 4\rangle$} &
			\multicolumn{1}{c|}{$\langle0, 0, 0, 0, 5, 3, 0, 0\rangle$} &
			\multicolumn{1}{c|}{$\langle0, 0, 0, 0, 5, 7, 0, 6\rangle$} &
			{$\langle0, 0, 0, 0, 0, 1, 5, 3\rangle$} \\ \hline
			
			\multicolumn{1}{|c|}{$\langle0, 0, 0, 0, 0, 3, 5, 9\rangle$} &
			\multicolumn{1}{c|}{$\langle0, 0, 0, 0, 0, 9, 5, 5\rangle$} &
			\multicolumn{1}{c|}{$\langle0, 0, 0, 0, 5, 3, 0, 1\rangle$} &
			{$\langle0, 0, 0, 0, 5, 7, 0, 7\rangle$} \\ \hline
			
			\multicolumn{1}{|c|}{$\langle0, 0, 0, 0, 0, 1, 5, 4\rangle$} &
			\multicolumn{1}{c|}{$\langle0, 0, 0, 0, 0, 7, 5, 0\rangle$} &
			\multicolumn{1}{c|}{$\langle0, 0, 0, 0, 0, 9, 5, 6\rangle$} &
			{$\langle0, 0, 0, 0, 5, 3, 0, 2\rangle$} \\ \hline
			
			\multicolumn{1}{|c|}{$\langle0, 0, 0, 0, 5, 7, 0, 8\rangle$} &
			\multicolumn{1}{c|}{$\langle0, 0, 0, 0, 0, 1, 5, 5\rangle$} &
			\multicolumn{1}{c|}{$\langle0, 0, 0, 0, 0, 7, 5, 1\rangle$} &
			{$\langle0, 0, 0, 0, 0, 9, 5, 7\rangle$} \\ \hline

			\multicolumn{1}{|c|}{$\langle0, 0, 0, 0, 5, 3, 0, 3\rangle$} &
			\multicolumn{1}{c|}{$\langle0, 0, 0, 0, 5, 7, 0, 9\rangle$} &
			\multicolumn{1}{c|}{$\langle0, 0, 0, 0, 0, 1, 5, 6\rangle$} &
			{$\langle0, 0, 0, 0, 0, 7, 5, 2\rangle$} \\ \hline
			
			\multicolumn{1}{|c|}{$\langle0, 0, 0, 0, 0, 9, 5, 8\rangle$} &
			\multicolumn{1}{c|}{$\langle0, 0, 0, 0, 5, 3, 0, 4\rangle$} &
			\multicolumn{1}{c|}{$\langle0, 0, 0, 0, 5, 9, 0, 0\rangle$} &
			{$\langle0, 0, 0, 0, 0, 1, 5, 7\rangle$}  \\ \hline

			\multicolumn{1}{|c|}{$\langle0, 0, 0, 0, 0, 7, 5, 3\rangle$} &
			\multicolumn{1}{c|}{$\langle0, 0, 0, 0, 0, 9, 5, 9\rangle$} &
			\multicolumn{1}{c|}{$\langle0, 0, 0, 0, 5, 3, 0, 5\rangle$} &
			{$\langle0, 0, 0, 0, 5, 9, 0, 1\rangle$}  \\ \hline
			
			\multicolumn{1}{|c|}{$\langle0, 0, 0, 0, 0, 1, 5, 8\rangle$} &
			\multicolumn{1}{c|}{$\langle0, 0, 0, 0, 0, 7, 5, 4\rangle$} &
			\multicolumn{1}{c|}{$\langle0, 0, 0, 0, 5, 1, 0, 0\rangle$} &
			{$\langle0, 0, 0, 0, 5, 3, 0, 6\rangle$}  \\ \hline
			
			\multicolumn{1}{|c|}{$\langle0, 0, 0, 0, 5, 9, 0, 2\rangle$} &
			\multicolumn{1}{c|}{$\langle0, 0, 0, 0, 0, 1, 5, 9\rangle$} &
			\multicolumn{1}{c|}{$\langle0, 0, 0, 0, 0, 7, 5, 5\rangle$} &
			{$\langle0, 0, 0, 0, 5, 1, 0, 1\rangle$}  \\ \hline
			
			\multicolumn{1}{|c|}{$\langle0, 0, 0, 0, 5, 3, 0, 7\rangle$} &
			\multicolumn{1}{c|}{$\langle0, 0, 0, 0, 5, 9, 0, 3\rangle$} &
			\multicolumn{1}{c|}{$\langle0, 0, 0, 0, 0, 3, 5, 0\rangle$} &
			{$\langle0, 0, 0, 0, 0, 7, 5, 6\rangle$} \\ \hline
			
			\multicolumn{1}{|c|}{$\langle0, 0, 0, 0, 5, 1, 0, 2\rangle$} &
			\multicolumn{1}{c|}{$\langle0, 0, 0, 0, 5, 3, 0, 8\rangle$} &
			\multicolumn{1}{c|}{$\langle0, 0, 0, 0, 5, 9, 0, 4\rangle$} &
			{$\langle0, 0, 0, 0, 0, 3, 5, 1\rangle$}  \\ \hline

			\multicolumn{1}{|c|}{$\langle0, 0, 0, 0, 0, 7, 5, 7\rangle$} &
			\multicolumn{1}{c|}{$\langle0, 0, 0, 0, 5, 1, 0, 3\rangle$} &
			\multicolumn{1}{c|}{$\langle0, 0, 0, 0, 5, 3, 0, 9\rangle$} &
			{$\langle0, 0, 0, 0, 5, 9, 0, 5\rangle$} \\ \hline
			
			\multicolumn{1}{|c|}{$\langle0, 0, 0, 0, 0, 3, 5, 2\rangle$} &
			\multicolumn{1}{c|}{$\langle0, 0, 0, 0, 0, 7, 5, 8\rangle$} &
			\multicolumn{1}{c|}{$\langle0, 0, 0, 0, 5, 1, 0, 4\rangle$} &
			{$\langle0, 0, 0, 0, 5, 7, 0, 0\rangle$}  \\ \hline
			
			\multicolumn{1}{|c|}{$\langle0, 0, 0, 0, 5, 9, 0, 6\rangle$} &
			\multicolumn{1}{c|}{$\langle0, 0, 0, 0, 0, 3, 5, 3\rangle$} &
			\multicolumn{1}{c|}{$\langle0, 0, 0, 0, 0, 7, 5, 9\rangle$} &
		    {$\langle0, 0, 0, 0, 5, 1, 0, 5\rangle$} \\ \hline
			
			\multicolumn{1}{|c|}{$\langle0, 0, 0, 0, 5, 7, 0, 1\rangle$} &
			\multicolumn{1}{c|}{$\langle0, 0, 0, 0, 5, 9, 0, 7\rangle$} &
			\multicolumn{1}{|c|}{$\langle0, 0, 0, 0, 0, 3, 5, 4\rangle$} &
		    {$\langle0, 0, 0, 0, 0, 9, 5, 0\rangle$} \\ \hline
			
			\multicolumn{1}{|c|}{$\langle0, 0, 0, 0, 5, 1, 0, 6\rangle$} &
			\multicolumn{1}{c|}{$\langle0, 0, 0, 0, 5, 7, 0, 2\rangle$} &
			\multicolumn{1}{c|}{$\langle0, 0, 0, 0, 5, 9, 0, 8\rangle$} &
		    {$\langle0, 0, 0, 0, 0, 3, 5, 5\rangle$}  \\ \hline
			
			\multicolumn{1}{|c|}{$\langle0, 0, 0, 0, 0, 9, 5, 1\rangle$} &
			\multicolumn{1}{c|}{$\langle0, 0, 0, 0, 5, 1, 0, 7\rangle$} &
			\multicolumn{1}{c|}{$\langle0, 0, 0, 0, 5, 7, 0, 3\rangle$} &
			{$\langle0, 0, 0, 0, 5, 9, 0, 9\rangle$} \\ \hline
		\end{tabular}
\end{table}

\item[--] \textbf{Third Filter}: We have identified 8 rules for clustering over the  second filter. These 8 rules is selected based on the examination of the cyclic structure with the minimum number of cycles in comparison to the other CAs.
All these rules demonstrate a maximum cycle length of $40$ for cell lengths 6 and 7, and $80$ for cell lengths 8 and 9, respectively, as outlined in Table~\ref{8 Rules}. Each rule is defined by its parameters and their respective values for different cell lengths ($n=6, 7, 8, 9$). The table provides details regarding the number of cycles and the maximum cycle length associated with each rule at various cell lengths.

\end{itemize}

\begin{table}[!ht]
	\centering
	\vspace{-1em}
	\caption {8 Selected Rules from Second filter 80 rules based on Chaotic Parameters}
	\label{8 Rules}
		\begin{tabular}{|l|l|l|l|l|l|l|}
			\hline
			FDCA Rules & $n=6$ & $n=7$ & Same For $n=6,7$ & $n=8$ & $n=9$ & Same For $n=8,9$ \\ \hline
			Rules & \#Cycles & \#Cycles & Max cycle length & \#Cycles & \#Cycles & Max cycle length \\ \hline
			$\langle 0, 0, 0, 0, 0, 1, 5, 1 \rangle$ & 25000 & 250000 & 40 & 1250000 & 12500000 & 80 \\ \hline
			$\langle0, 0, 0, 0, 0, 1, 5, 3\rangle$ & 25000 & 250000 & 40 & 1250000 & 12500000 & 80 \\ \hline
			$\langle0, 0, 0, 0, 0, 1, 5, 7\rangle$ & 25000 & 250000 & 40 & 1250000 & 12500000 & 80 \\ \hline
			$\langle0, 0, 0, 0, 0, 1, 5, 9\rangle$ & 25000 & 250000 & 40 & 1250000 & 12500000 & 80 \\ \hline
			$\langle0, 0, 0, 0, 5, 1, 0, 1\rangle$ & 25000 & 250000 & 40 & 1250000 & 12500000 & 80 \\ \hline
			$\langle0, 0, 0, 0, 5, 1, 0, 3\rangle$ & 25000 & 250000 & 40 & 1250000 & 12500000 & 80 \\ \hline
			$\langle0, 0, 0, 0, 5, 1, 0, 7\rangle$ & 25000 & 250000 & 40 & 1250000 & 12500000 & 80 \\ \hline
			$\langle0, 0, 0, 0, 5, 1, 0, 9\rangle$ & 25000 & 250000 & 40 & 1250000 & 12500000 & 80 \\ \hline
		\end{tabular}
	\vspace{-1em}
	
\end{table}

\subsection{Final List of Rules}
Out of a total of 1560 reversible rules, our selection process narrowed down the candidates for clustering to a subset of rules meeting specific criteria. From the criteria related to the cyclic structure of the rules (Section~\ref{criteria I: Cyclic structure of the rules}), we identify 28 rules. Additionally, from the criteria based on chaotic parameters (Section~\ref{criteria II: Chaotic parameter}), we identify 8 rules. In total, we have 36 selected reversible FDCA rules considered as candidates for clustering purposes. Table~\ref{36 Rule} presents these 36 selected reversible FDCA rules. These rules were chosen based on their potential to contribute to our clustering analysis. They represent a subset of the total reversible rules that exhibit promising properties for our clustering objectives. These selected rules will be employed for clustering analysis in the subsequent stages of the study.

\begin{table}[hbtp]
	\caption{Final List of 36 Rule for Clustering using Decimal FDCA}
	\label{36 Rule}
	\begin{tabular}{|cccc|}
		\hline
		\multicolumn{4}{|c|}{\textbf{Final Selected 36 Rule From Filtering Criteria I \& II}}                                                                                                                                                                                                                                                                                     \\ \hline
		\multicolumn{1}{|c|}{$\langle0,0,0,0,1,7,8,1\rangle$} & \multicolumn{1}{c|}{$\langle0,0,0,0,3,7,6,7\rangle$} & \multicolumn{1}{c|}{$\langle0,0,0,0,4,9,3,1\rangle$} & 
		{$\langle0,0,0,0,6,7,3,7\rangle$}  \\ \hline
		
		\multicolumn{1}{|c|}{$\langle0,0,0,0,3,9,4,1\rangle$} & 
		\multicolumn{1}{c|}{$\langle0,0,0,0,0,1,5,7\rangle$} &
		\multicolumn{1}{c|}{$\langle0,0,0,0,1,7,8,3\rangle$} & 
		{$\langle0,0,0,0,3,7,6,9\rangle$} \\ \hline
		
		 \multicolumn{1}{|c|}{$\langle0,0,0,0,4,9,3,3\rangle$} & \multicolumn{1}{c|}{$\langle0,0,0,0,6,7,3,9\rangle$} & \multicolumn{1}{c|}{$\langle0,0,0,0,3,9,4,3\rangle$} & 
		 {$\langle0,0,0,0,0,1,5,9\rangle$} \\ \hline
		 
		\multicolumn{1}{|c|}{$\langle0,0,0,0,1,7,8,7\rangle$} & \multicolumn{1}{c|}{$\langle0,0,0,0,3,9,4,1\rangle$} & \multicolumn{1}{c|}{$\langle0,0,0,0,4,9,3,7\rangle$} & 
		{$\langle0,0,0,0,3,7,6,1\rangle$} \\ \hline
		
		 \multicolumn{1}{|c|}{$\langle0,0,0,0,3,9,4,7\rangle$} & 
		 \multicolumn{1}{c|}{$\langle0,0,0,0,5,1,0,1\rangle$} &
	   	 \multicolumn{1}{c|}{$\langle0,0,0,0,1,7,8,9\rangle$} & 
		 {$\langle0,0,0,0,3,9,4,3\rangle$} \\ \hline
		
		 \multicolumn{1}{|c|}{$\langle0,0,0,0,4,9,3,9\rangle$} & \multicolumn{1}{c|}{$\langle0,0,0,0,3,7,6,3\rangle$} & \multicolumn{1}{c|}{$\langle0,0,0,0,3,9,4,9\rangle$} & 
		 {$\langle0,0,0,0,5,1,0,3\rangle$} \\ \hline
		 
		\multicolumn{1}{|c|}{$\langle0,0,0,0,3,7,6,1\rangle$} & \multicolumn{1}{c|}{$\langle0,0,0,0,3,9,4,7\rangle$} & \multicolumn{1}{c|}{$\langle0,0,0,0,6,7,3,1\rangle$} & 
		{$\langle0,0,0,0,3,7,6,7\rangle$} \\ \hline
		
		 \multicolumn{1}{|c|}{$\langle0,0,0,0,0,1,5,1\rangle$} & 
		 \multicolumn{1}{c|}{$\langle0,0,0,0,5,1,0,7\rangle$} &
		\multicolumn{1}{c|}{$\langle0,0,0,0,3,7,6,3\rangle$} &  
		{$\langle0,0,0,0,3,9,4,9\rangle$}  \\ \hline
		
		 \multicolumn{1}{|c|}{$\langle0,0,0,0,6,7,3,3\rangle$} & \multicolumn{1}{c|}{$\langle0,0,0,0,3,7,6,9\rangle$} & \multicolumn{1}{c|}{$\langle0,0,0,0,0,1,5,3\rangle$} & 
		 {$\langle0,0,0,0,5,1,0,9\rangle$} \\ \hline
	\end{tabular}
\vspace{-1em}
\end{table}

\section{Clustering Technique}
For a high dimensional datasets, the G{\"o}del encoded string may become substantially large, so employing the CA based clustering algorithm on it can lead to increased computation time due to the necessity of exploring all configurations. However, by partitioning the G{\"o}del encoded strings into several segments such that each segment is processed independently until the length of string is reduced, we can significantly mitigate the time complexity. This approach allows us the flexibility to apply distinct rules for different segments of the same object. Hence, we opt for the \emph{Vertical Splitting} method, which is detailed below to illustrate its application.\\

\textbf{Vertical Splitting:} As illustrated in Figure~\ref{fig:godel}, the G{\"o}del numbers generated per each row of the dataset undergoes a process known as Vertical Splitting, where the G{\"o}del numbers treated as decimal string are partitioned into multiple splits based on the split size. This splitting is implemented on the dataframe, ensuring that these segments share the same length except the last segment which can be of smaller length. In cases where the lengths of the dataframes differ, zero-padding is applied at the beginning of that dataframe. 

\begin{figure}[h]
	\vspace{-1em}
	\centering
	\includegraphics[scale=0.45]{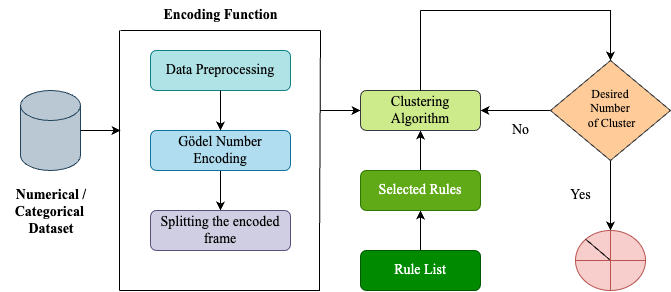}
	\caption{Flow diagram of the Implementation} \label{fig5}
\end{figure}

Directly applying the clustering algorithm to the entire dataframe might result in extended processing times. However, by dividing the dataframe into multiple splits, each part can be processed in parallel. Initially, the clustering algorithm may not yield the desired number of clusters. To attain the desired cluster count, clusters exceeding this number are merged. This subsequent merging of clusters, based on a specified metric, is crucial to achieving the desired number of clusters. Figure \ref{fig5} is the flow diagram of the implementation. Datasets are encoded using an encoding function and then split into numbers of data frames. Decimal FDCA based clustering algorithm is applied to that dataframe and results are achieved. In the next section we discuss three metrics based on  Silhouette score, average/Median, and maximum degree of participation which are employed for cluster merging decisions.

\subsection{Silhouette Score based Metric}
In the study conducted by Ref.~\cite{abhishek2023cellular, manoranjan2023optimized}, the Silhouette score serves as a pivotal metric for merging the smallest cluster with others. The approach begins by identifying the cluster with the fewest elements and systematically incorporates this cluster into others. Throughout this process, the silhouette score is continuously computed. The merging continues iteratively, with each step involving the addition of elements from the smallest cluster to the cluster that yields the highest silhouette score. This process is repeated until the desired number of clusters is achieved. In this paper, we apply the following process as described with an example:

\begin{enumerate}
	\item \textbf{Iterate through clusters and find the cluster with the least elements:}\\
	Let us say we have four clusters: Cluster A (5 elements), Cluster B (7 elements), Cluster C (4 elements), and Cluster D (6 elements). Among these, Cluster C has the least elements.
	\item \textbf{Pick the least elements cluster, and add them to all other clusters individually and calculate the silhouette score:}
	\begin{itemize}
		\item Add Cluster C to Cluster A, calculate silhouette score.
		\item Add Cluster C to Cluster B, calculate silhouette score.
		\item Add Cluster C to Cluster D, calculate silhouette score.
	\end{itemize}
	\item \textbf{Merge the least element cluster with the cluster that has the maximum silhouette score:}\\
	Suppose Cluster C has the maximum silhouette score when merged with Cluster B, then, merge Cluster C with Cluster B.
	\item \textbf{Repeat Step $1$ to $3$ until the desired number of clusters is achieved:}\\
	Now, we have three clusters: Cluster A (5 elements), Cluster BC (11 elements), and Cluster D (6 elements). Let the desired number of clusters is two. Then, repeat the process: find the cluster with the least elements (Cluster A), add it to each of all other clusters, calculate silhouette scores, and merge it with the cluster that yields the maximum silhouette score. Continue this process until the desired number of clusters is reached.
   \end{enumerate}

In this iterative process, we continuously merge clusters based on silhouette scores, gradually forming bigger clusters until we achieve the desired number. This method helps optimize cluster formation by considering both the size of clusters and their silhouette scores, ensuring a balanced and well-separated clustering solution. However, the problem of this approach is its dependency on the benchmark validation index to form clusters and the resulting time complexity.

\subsection{Merging Metric based on Average}
\begin{figure}[!h]
	\centering
	\includegraphics[width=12cm,height=6cm]{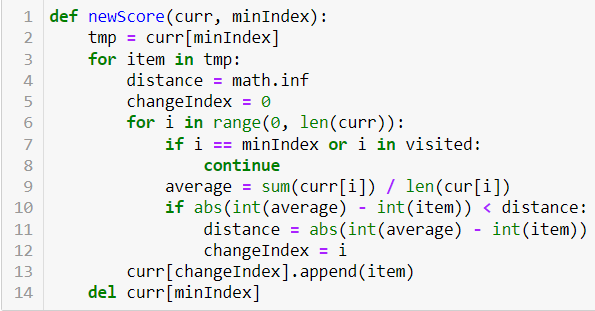}
	\caption{Pseudo-Code to merge clusters using average metric} 
	\label{Avg metric code}
\end{figure}

We have developed a framework utilizing G{\"o}del number based encoding for our data frames. As these G{\"o}del number are natural numbers, we can exploit arithmetic properties to merge the clusters. Our approach focuses on minimizing intra-cluster distance when integrating new elements, thereby enhancing the likelihood of their inclusion in the cluster. To ensure this, we define a new distance metric based on average of the cluster elements. This distance computation involves averaging G{\"o}del numbers of configurations (data elements) within each cluster and determining the closest cluster for an element to be added such that the change in cluster average is minimal after this addition. The code in Figure \ref{Avg metric code} demonstrates how to merge the selected cluster with another cluster that yields a smaller distance to the original cluster. Additionally, instead of the average, the median can also be employed for distance assessment giving similar result.

\begin{figure}[!h]
	\centering
	\vspace{-1em}
	\includegraphics[scale=0.4]{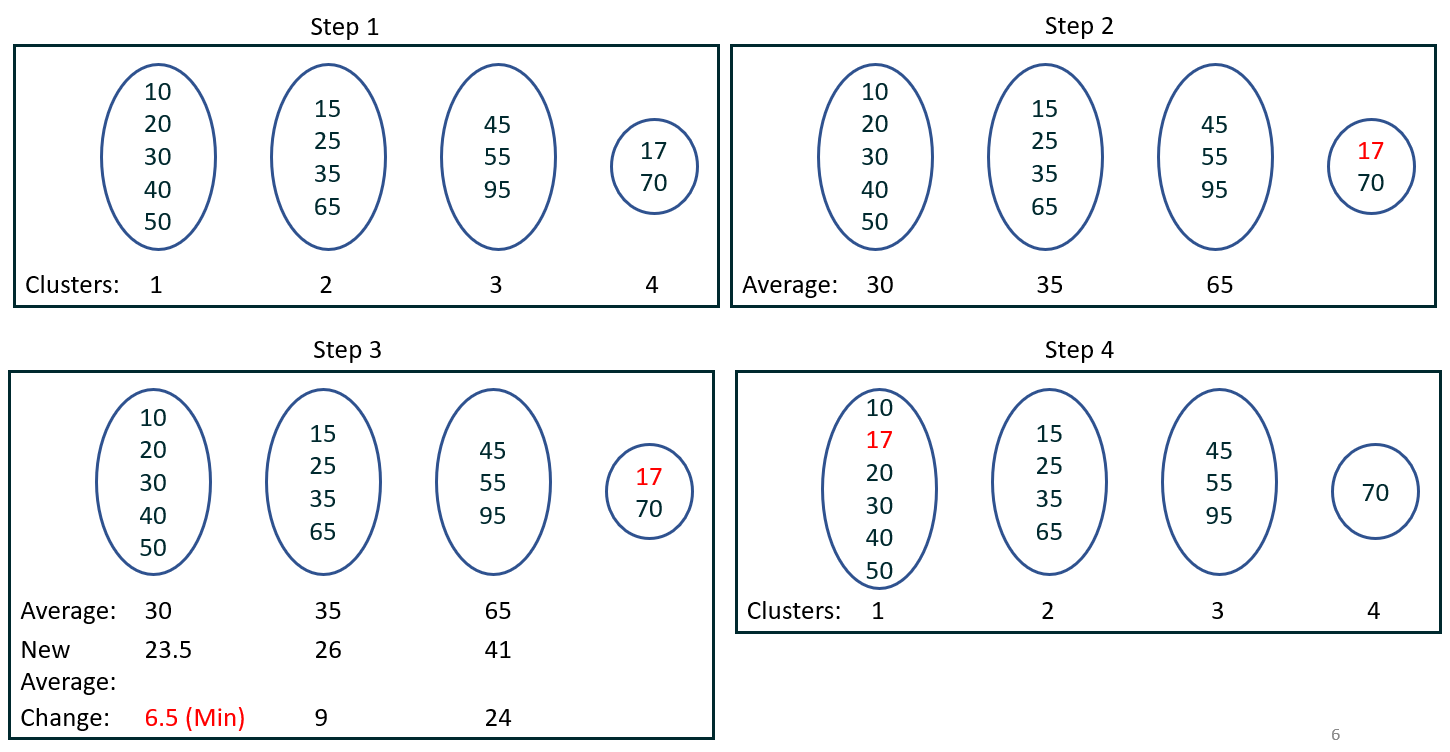}
\caption{Example of merging the clusters using average metric} 
	\label{fig14}
	\vspace{-1em}
\end{figure} 

For instance, in Figure \ref{fig14}, we observe four clusters with Cluster 4 having the fewest elements. In Step 2, the average for each cluster is calculated. Moving to Step 3,  we select the smallest cluster, and choose one element, (say, element 17). Then, we introduce element 17 into each cluster and track the change in the average for each cluster.
The objective is to identify the cluster where the addition of element 17 results in the minimum change in the average.
This minimum change indicates a higher likelihood that element 17 belongs to that cluster. Consequently, element 17 is assigned to Cluster 1 (Step 4). This process iterates until all elements are appropriately assigned to desired number of clusters based on minimizing changes in cluster averages.

\subsection{Maximum Degree of Participation based metric}
To facilitate the merging of clusters from one vertical split to another, we adopt an approach similar to Ref.~\cite{mukherjee2021clustering,mukherjee2021reversible}, which involves calculating the \emph{degree of participation} or \emph{participation score}. In Ref.~\cite{mukherjee2021clustering,mukherjee2021reversible} this participation score was calculated to merge clusters from different spatial levels. However, here, our approach enhances the process by leveraging the maximum degree of participation of the data sub-elements for each split and subsequently merging the vertical clusters. In our algorithm, the dataset undergoes a vertical split into several parts, and each split is independently clustered based on certain FDCA rules. These clusters for each split is called the \emph{primary clusters}. In the second stage, the goal is to amalgamate these primary clusters to establish a clustering over the original dataset. During this merging process, we use a different CA which can potentially yield improved results, as it introduces a distinct bijective mapping, contributing to a more refined clustering outcome. This new CA is called an \emph{auxiliary CA}, and based on this, a participation score is calculated. 

To use this metric, the primary condition is, each vertical split is of same size, say $l$, such that all the sub-configurations are of length $l$. Now apply the following steps:
\begin{enumerate}
	\item Take the union of all sub-configurations for each split; let that set be called the \emph{auxiliary configurations}.
	\item Randomly choose a new CA rule from the potential list of candidate rules. This CA is the auxiliary CA. 
	\item Apply this auxiliary CA over the auxiliary configurations and make clusters. Let these clusters be named as \emph{auxiliary clusters}.
	\item With respect to these auxiliary clusters, calculate the degree of participation of all primary clusters for each split and tabulate.
	
		\textbf{Degree of Participation:} Let $c_{i,j}$ be the $j^{th}$ primary cluster of Split $i$ where $|c_{i,j}| = v_j$, and $C_t$ be an auxiliary cluster. Let $v'_j$ be the number of (sub-)configurations from primary cluster $c_{i,j}$ in the auxiliary cluster $C_t$. The degree of participation is represented by $\mu(C_t, c_{i,j}) = \frac{v'_j}{v_j}$.
	
	\item \textbf{Merging:} For each auxiliary cluster, choose those primary clusters from each split for which the degree of participation is maximum and more than $50\%$. Merge the original data elements corresponding to these clusters into a new \emph{temporary} cluster. Continue this step until all auxiliary clusters are considered.
	
	If number of desired clusters is less than the number of temporary clusters formed and at least $50\%$ of the data elements of one temporary cluster matches with another cluster, merge these temporary cluster to form a new cluster. Otherwise, remove the common elements from the larger cluster and keep the other temporary clusters as found.
	
	\item Report the temporary clusters as final clusters after merging based on this maximum participation score. 
\end{enumerate} 
\begin{figure}[!h]
	\centering
	\vspace{-1em}
	\includegraphics[scale=0.5]{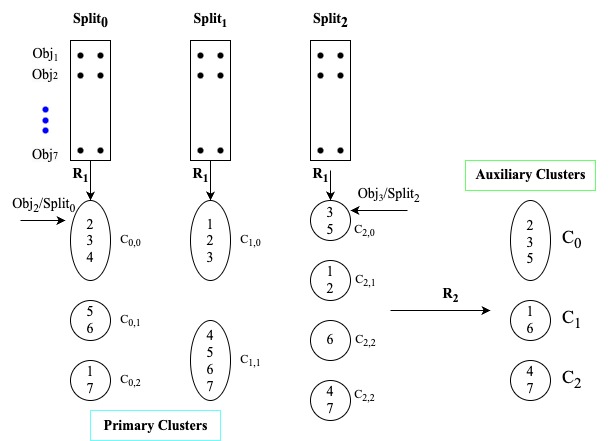}
	\caption{Formation of Auxiliary Clusters} 
	\label{Fig:Maximum Participation} 
	\vspace{-1em}
\end{figure}

For example, consider a hypothetical Figure~\ref{Fig:Maximum Participation}, where we have 7 data points split into 3 equal parts and clustered using a CA rule $R_1$. Let the primary clusters of the first split be $c_{0,0}, c_{0,1}, c_{0,2}$, second split be $c_{1,0}, c_{1,1}$ and the third split be $c_{2,0}, c_{2,1}, c_{2,2}, c_{2,3}$. Taking a union of all data sub-points of each split, we get a new set of auxiliary configurations. Using an auxiliary CA $R_2$, these data points are divided into 3 clusters $C_0, C_1$ and $C_2$. Table \ref{Tab:participation_score} depicts the degree of participation in percentage of each primary cluster with respect to the auxiliary clusters. 

%
%

\begin{table}[hbtp]
	\centering
	\vspace{-1em}
	\caption{Degree of Participation of the Primary Clusters with respect to the Auxiliary Clusters}
	\label{Tab:participation_score}
		\begin{tabular}{|l|cc|cc|cc|}
			\hline
			\multicolumn{1}{|c|}{\textbf{Auxiliary Cluster}} & \multicolumn{2}{c|}{\textbf{$Split0$}}                               & \multicolumn{2}{c|}{\textbf{$Split1$}}                               & \multicolumn{2}{c|}{\textbf{$Split2$}} \\ \hline
			\multirow{4}{*}{$C_0$}                            & \multicolumn{1}{c|}{$c_{0,0}$}                  & 66\%                  & \multicolumn{1}{c|}{$c_{1,0}$}                  & 66\%                  & \multicolumn{1}{c|}{$c_{2,0}$}   & 100\%  \\ 
			\hhline{~------} 
			& \multicolumn{1}{c|}{$c_{0,1}$}                  & 50\%                  & \multicolumn{1}{c|}{\multirow{3}{*}{$c_{1,1}$}} & \multirow{3}{*}{25\%} & \multicolumn{1}{c|}{$c_{2,1}$}   & 50\%   \\ 
			\hhline{~--~~--} 
			& \multicolumn{1}{c|}{\multirow{2}{*}{$c_{0,2}$}} & \multirow{2}{*}{0\%}  & \multicolumn{1}{c|}{}                        &                       & \multicolumn{1}{c|}{$c_{2,2}$}   & 0\%    \\ \hhline{~~~~~--} 
			& \multicolumn{1}{c|}{}                        &                       & \multicolumn{1}{c|}{}                        &                       & \multicolumn{1}{c|}{$c_{2,3}$}   & 0\%    \\ \hline
			\multirow{4}{*}{$C_1$}                            & \multicolumn{1}{c|}{$c_{0,0}$}                  & 0\%                   & \multicolumn{1}{c|}{$c_{1,0}$}                  & 33\%                  & \multicolumn{1}{c|}{$c_{2,0}$}   & 0\%    \\ 
			\hhline{~------} 
			& \multicolumn{1}{c|}{$c_{0,1}$}                  & 50\%                  & \multicolumn{1}{c|}{\multirow{3}{*}{$c_{1,1}$}} & \multirow{3}{*}{25\%} & \multicolumn{1}{c|}{$c_{2,1}$}   & 50\%   \\ 			\hhline{~--~~--} 
			& \multicolumn{1}{c|}{\multirow{2}{*}{$c_{0,2}$}} & \multirow{2}{*}{50\%} & \multicolumn{1}{c|}{}                        &                       & \multicolumn{1}{c|}{$c_{2,2}$}   & 100\%  \\ \hhline{~~~~~--}  
			& \multicolumn{1}{c|}{}                        &                       & \multicolumn{1}{c|}{}                        &                       & \multicolumn{1}{c|}{$c_{2,3}$}   & 0\%    \\ \hline
			\multirow{4}{*}{$C_2$}                            & \multicolumn{1}{c|}{$c_{0,0}$}                  & 33\%                  & \multicolumn{1}{c|}{$c_{1,0}$}                  & 0\%                   & \multicolumn{1}{c|}{$c_{2,0}$}   & 0\%    \\ \hhline{~------}
			& \multicolumn{1}{c|}{$c_{0,1}$}                  & 0\%                   & \multicolumn{1}{c|}{\multirow{3}{*}{$c_{1,1}$}} & \multirow{3}{*}{50\%} & \multicolumn{1}{c|}{$c_{2,1}$}   & 0\%    \\ 	\hhline{~--~~--} 
			& \multicolumn{1}{c|}{\multirow{2}{*}{$c_{0,2}$}} & \multirow{2}{*}{50\%} & \multicolumn{1}{c|}{}                        &                       & \multicolumn{1}{c|}{$c_{2,2}$}   & 0\%    \\ \hhline{~~~~~--}  
			& \multicolumn{1}{c|}{}                        &                       & \multicolumn{1}{c|}{}                        &                       & \multicolumn{1}{c|}{$c_{2,3}$}   & 100\%  \\ \hline
		\end{tabular}
	\vspace{-1em}
\end{table}

For instance, if we consider cluster $c_{0,0}$, among the 3 data objects ${2,3,4}$, only ${2,3}$ are present in the auxiliary cluster $C_0$. Therefore, the participation score $\mu(C_0, c_{0,0})$ is $\frac{2}{3} = 67\%$. Similarly, if all elements of a primary cluster is present in the auxiliary cluster, then maximum participation score is $100\%$, and if no data objects are common, then the minimum participation score is $0\%$.

We aim to merge the data objects that have the maximum participation score from each partition. Therefore, we merge $c_{0,0}, c_{1,0}$ and $c_{2,0}$, with respect to $C_0$ as their degree of participation is more than 50\%. This forms a new temporary cluster of data elements \{1,2,3,4,5\}. Next, we take $c_{2,3}$ with respect to $C_2$ and get a cluster \{4,7\}. We also have element 6 forming a new cluster as for it everywhere the participation score is 50\%. Now, if our target is to get two clusters, we can merge the temporary clusters \{1,2,3,4,5\} and \{4,7\} to form a new cluster \{1,2,3,4,5,7\} along with the other cluster \{6\}. Else, we can report three clusters as:
\begin{enumerate}[label=\alph*), noitemsep]
	\item Cluster 1: $\{1, 2, 3, 5\}$
	\item Cluster 2: $\{4, 7\}$
	\item Cluster 3: $\{6\}$
\end{enumerate}


\subsection{The clustering Algorithm by Reversible Decimal FDCA} In this section, we discuss our overall clustering algorithm step by step.
\subsubsection*{Stage 1: Dataframe Generation based on Gödel Numbers}
\begin{enumerate}[label=\textbf{Step \arabic*:}]
	\label{algorithm}
	\item Generate the dataframe from the original dataset based on Gödel Numbers.
	\item Divide the dataframe into a number of splits where the size of each split can vary from $6$ to $9$ (6 is preferred). If the choice of metric is maximum degree of participation, the dataframe size need to be a multiple of the split size. To have this, zeros may have to be added at the beginning of the dataframes.
	\item Go to Stage 2.
\end{enumerate}

\subsubsection*{Stage 2: Apply the Rule}
\begin{enumerate}[label=\textbf{Step \arabic*:}]
	\item Select a random rule from a set of candidate rules and apply it in the next step.
	\item For each split, apply the same rule over the configurations of that split until all configurations are clustered. (Same rule is chosen as using the same rule for each split yields better average scores.)
	\item If split size is equal, current number of clusters is more than the desired number, and maximum participation score is chosen as metric, go to Stage 3, Option 3. \\ Otherwise, encode the clusters resulting from the previous step using decimal numbers:
	\begin{enumerate}[label=\textbf{\alph*)}]
		\item Find the number of clusters for each split.
		\item If the number of clusters for the $i^{th}$ split is $m_i$, encode each cluster using $\lceil \log_{10} m_i \rceil$ digits.
		\item The (sub-)configuration that are in the same cluster are encoded by the same digits.
		\item Concatenate the encoded strings of each split to have new data frames corresponding to each object of the dataset.
		\item Go to Stage 3
	\end{enumerate}
	
\end{enumerate}

\subsubsection*{Stage 3: Iterative Algorithm to get Desired Number of Clusters}
\begin{enumerate}[label=\textbf{Case \arabic*:}]
	\item Current number of clusters is more than $d$ (desired number of clusters)
	\begin{enumerate}[label=\textbf{Option \arabic*:}]
		\item Silhouette Score
		\begin{enumerate}[label=\textbf{\arabic*.}]
			\item Iterate through clusters and find the cluster with the least elements.
			\item Add the least elements cluster to all other clusters and calculate the silhouette score.
			\item Merge the least element cluster with the cluster that has the maximum silhouette score.
			\item Repeat until the desired number of clusters is achieved.
			\item Exit
		\end{enumerate}
		\item Average Metric
		\begin{enumerate}[label=\textbf{\arabic*.}]
			\item Iterate through clusters and find the cluster with the least elements.
			\item Calculate the average value of the cluster for all clusters except the cluster with the least element.
			\item For each element in the cluster with the least elements, pick each element and calculate the distance between the current element and the average of the clusters.
			\item Add the chosen element to the cluster which has the minimum distance.
			\item Repeat until the desired number of clusters is achieved.
			\item Exit
		\end{enumerate}
		\item Maximum Participation Score
		\begin{enumerate}[label=\textbf{\arabic*.}]
			\item Randomly select a new (auxiliary CA) rule and generate the auxiliary clusters.
			\item Create a table for maximum participation score with the auxiliary clusters and merge the clusters according to the maximum participation score percentage.
			\item Repeat the process (considering only one vertical split for the new clusters generated in previous step) until the desired number of clusters is achieved.
		\end{enumerate}
	\end{enumerate}
	
	\item Current number of clusters is less than $d$
	\begin{enumerate}[label=\textbf{Step \arabic*:}]
		\item Set $k = 1$ and select a new (auxiliary CA) rule from the rule set.
		\item Apply the rule and generate auxiliary clusters. Suppose the number of clusters is $e$.
		\item If $e = d$, return the output and exit the algorithm. Else, go to Step 4.
		\item Find the elements in each cluster that belong to the same auxiliary cluster, and group them together to form a new cluster containing only those elements. Set $k = k + 1$.
		\item Repeat Step 4 until all data elements are clustered.
		\item If \(k < d\), go to Step 1 and repeat.

	\end{enumerate}
\end{enumerate}

Interestingly, we find that utilizing different FDCA rules in different stages often yields better scores compared to using the same rule. This observation suggests that varying FDCA rule can significantly impact the clustering results, potentially leading to improved performance metrics.

\section{Results}
In this study, we explore the utilization of clustering techniques by decimal FDCA across a range of datasets. Our findings showcase the efficacy of these techniques in effectively identifying clusters within the data. 
These results demonstrate the effectiveness of different clustering methods in partitioning the dataset. These findings suggest that the choice of rule and merging method significantly impacts the clustering performance, with variations observed across different evaluation metrics.

\begin{table}[!h]
	\centering
	\caption{ Comparison of Merging Method based on evaluation metrics for the Seeds and User Knowledge Modelling Dataset}
	\label{FDCA Rules with Proposed Algorithm-1}
	\begin{tabular}{|c|cccc|}
		\hline
		\multirow{3}{*}{\begin{tabular}[c]{@{}c@{}}\textbf{PROPOSED}\\ \textbf{ALGORITHM}\end{tabular}} &
		\multicolumn{4}{c|}{\textbf{Dataset Name}} \\ \hhline{~----} 
		&
		\multicolumn{4}{c|}{\textbf{Seeds}}  \\ \hhline{~----} 
		&
		\multicolumn{1}{c|}{\textbf{Silhouette}} &
		\multicolumn{1}{c|}{\textbf{DB}} &
		\multicolumn{1}{c|}{\textbf{CH}} &
		\textbf{Dunn Index} \\ \hline
		\begin{tabular}[c]{@{}c@{}}Silhouette Score\\ metric\end{tabular} &
		\multicolumn{1}{c|}{0.714} &
		\multicolumn{1}{c|}{10.113} &
		\multicolumn{1}{c|}{688.135} &
		0.0052 \\ \hline
		FDCA Rules &
		\multicolumn{1}{c|}{$\langle0,0,0,0,0,1,5,7\rangle$} &
		\multicolumn{1}{c|}{$\langle0,0,0,0,3,9,4,9\rangle$} &
		\multicolumn{1}{c|}{$\langle0,0,0,0,1,7,8,7\rangle$} &
		{$\langle0,0,0,0,1,7,8,3\rangle$} \\ \hline
		Average Metric &
		\multicolumn{1}{c|}{0.529} &
		\multicolumn{1}{c|}{10.231} &
		\multicolumn{1}{c|}{680.914} &
		0.0030 \\ \hline
		FDCA Rules &
		\multicolumn{1}{c|}{$\langle0,0,0,0,3,9,4,7\rangle$} &
		\multicolumn{1}{c|}{$\langle0,0,0,0,6,7,3,9\rangle$} &
		\multicolumn{1}{c|}{$\langle0,0,0,0,1,7,8,7\rangle$} &
		\begin{tabular}[c]{@{}c@{}}{$\langle0,0,0,0,1,7,8,1\rangle$}\\ and\\ {$\langle0,0,0,0,3,7,6,3\rangle$}\end{tabular} \\ \hline
		Maximum participation &
		\multicolumn{1}{c|}{0.795} &
		\multicolumn{1}{c|}{10.086} &
		\multicolumn{1}{c|}{685.351} &
		0.0032 \\ \hline
		FDCA Rules &
		\multicolumn{1}{c|}{$\langle0,0,0,0,0,1,5,7\rangle$} &
		\multicolumn{1}{c|}{$\langle0,0,0,0,3,7,6,7\rangle$}&
		\multicolumn{1}{c|}{$\langle0,0,0,0,3,7,6,7\rangle$} &
		{$\langle0,0,0,0,1,7,8,1\rangle$} \\ \hline  \hline
		\multirow{3}{*}{\begin{tabular}[c]{@{}c@{}}\textbf{PROPOSED}\\ \textbf{ALGORITHM}\end{tabular}} &
		\multicolumn{4}{c|}{\textbf{Dataset Name}}  \\ \hhline{~----} 
		&
		\multicolumn{4}{c|}{\textbf{User Knowledge Modelling}} \\ \hhline{~----} 
		&
		\multicolumn{1}{c|}{\textbf{Silhouette}} &
		\multicolumn{1}{c|}{\textbf{DB}} &
		\multicolumn{1}{c|}{\textbf{CH}} &
		\textbf{Dunn Index} \\ \hline
		\begin{tabular}[c]{@{}c@{}}Silhouette Score\\ metric\end{tabular} &
		\multicolumn{1}{c|}{0.783} &
		\multicolumn{1}{c|}{0.628} &
		\multicolumn{1}{c|}{249.447} &
		0.0080 \\ \hline
		FDCA Rules &
		\multicolumn{1}{c|}{$\langle0,0,0,0,1,7,8,1\rangle$} &
		\multicolumn{1}{c|}{$\langle0,0,0,0,3,7,6,9\rangle$} &
		\multicolumn{1}{c|}{$\langle0,0,0,0,3,9,4,1\rangle$}&
		{$\langle0,0,0,0,6,7,3,7\rangle$} \\ \hline
		Average Metric &
		\multicolumn{1}{c|}{0.756} &
		\multicolumn{1}{c|}{0.609} &
		\multicolumn{1}{c|}{247.675} &
		0.0082 \\ \hline
		FDCA Rules &
		\multicolumn{1}{c|}{$\langle0,0,0,0,1,7,8,9\rangle$} &
		\multicolumn{1}{c|}{$\langle0,0,0,0,1,7,8,9\rangle$} &
		\multicolumn{1}{c|}{$\langle0,0,0,0,3,9,4,9\rangle$} &
		{$\langle0,0,0,0,6,7,3,9\rangle$} \\ \hline
		Maximum participation &
		\multicolumn{1}{c|}{0.824} &
		\multicolumn{1}{c|}{0.808} &
		\multicolumn{1}{c|}{449.141} &
		0.0107 \\ \hline
		FDCA Rules &
		\multicolumn{1}{c|}{$\langle0,0,0,0,3,7,6,3\rangle$} &
		\multicolumn{1}{c|}{\begin{tabular}{c}$\langle 0,0,0,0,3,7,6,7 \rangle$ \\ and \\  $\langle 0,0,0,0,5,1,0,3 \rangle$\end{tabular}} &
		\multicolumn{1}{c|}{$\langle0,0,0,0,3,9,4,9\rangle$} &
		{$\langle0,0,0,0,4,9,3,9\rangle$}\\ \hline
	\end{tabular}
\end{table}

\begin{table}[!h]
	\centering
	\caption{Comparison of Merging Method based on evaluation metrics for the Heart Failure Clinical and Buddy Move Dataset
	}
	\label{FDCA Rules with Proposed Algorithm-2}
	\begin{tabular}{|c|cccc|}
		\hline
		\multirow{3}{*}{\begin{tabular}[c]{@{}c@{}}\textbf{PROPOSED}\\ \textbf{ALGORITHM}\end{tabular}} &
		\multicolumn{4}{c|}{\textbf{Dataset Name}} \\ \hhline{~----} 
		&
		\multicolumn{4}{c|}{\textbf{Heart Failure Clinical}}  \\ \hhline{~----} 
		&
		\multicolumn{1}{c|}{\textbf{Silhouette}} &
		\multicolumn{1}{c|}{\textbf{DB}} &
		\multicolumn{1}{c|}{\textbf{CH}} &
		\textbf{Dunn Index} \\ \hline
		\begin{tabular}[c]{@{}c@{}}Silhouette Score\\ metric\end{tabular} &
		\multicolumn{1}{c|}{0.619} &
		\multicolumn{1}{c|}{1.571} &
		\multicolumn{1}{c|}{837.147} &
		0.0107 \\ \hline
		FDCA Rules &
		\multicolumn{1}{c|}{$\langle0,0,0,0,3,7,6,7\rangle$}  &
		\multicolumn{1}{c|}{$\langle0,0,0,0,4,9,3,1\rangle$} &
		\multicolumn{1}{c|}{$\langle0,0,0,0,1,7,8,9\rangle$} &
		{$\langle0,0,0,0,3,7,6,1\rangle$}  \\ \hline
		Average Metric &
		\multicolumn{1}{c|}{0.587} &
		\multicolumn{1}{c|}{0.651} &
		\multicolumn{1}{c|}{248.191} &
		0.0107 \\ \hline
		FDCA Rules &
		\multicolumn{1}{c|}{$\langle0,0,0,0,1,7,8,1\rangle$}  &
		\multicolumn{1}{c|}{$\langle0,0,0,0,5,1,0,3\rangle$} &
		\multicolumn{1}{c|}{$\langle0,0,0,0,5,1,0,1\rangle$}  &
		{$\langle0,0,0,0,4,9,3,1\rangle$}  \\ \hline
		Maximum participation &
		\multicolumn{1}{c|}{0.728} &
		\multicolumn{1}{c|}{1.604} &
		\multicolumn{1}{c|}{822.47} &
		0.0108 \\ \hline
		FDCA Rules &
		\multicolumn{1}{c|}{$\langle0,0,0,0,3,7,6,3\rangle$}  &
		\multicolumn{1}{c|}{$\langle0,0,0,0,0,1,5,9\rangle$} &
		\multicolumn{1}{c|}{$\langle0,0,0,0,4,9,3,7\rangle$}  &
		{$\langle0,0,0,0,3,9,4,9\rangle$}  \\ \hline \hline
		\multirow{3}{*}{\begin{tabular}[c]{@{}c@{}}\textbf{PROPOSED}\\ \textbf{ALGORITHM}\end{tabular}} &
		\multicolumn{4}{c|}{\textbf{Dataset Name}} \\ \hhline{~----} 
		&
		\multicolumn{4}{c|}{\textbf{Buddy Move}} \\ \hhline{~----} 
		&
		\multicolumn{1}{c|}{\textbf{Silhouette}} &
		\multicolumn{1}{c|}{\textbf{DB}} &
		\multicolumn{1}{c|}{\textbf{CH}} &
		\textbf{Dunn Index} \\ \hline
		\begin{tabular}[c]{@{}c@{}}Silhouette Score\\ metric\end{tabular} &
		\multicolumn{1}{c|}{0.536} &
		\multicolumn{1}{c|}{0.782} &
		\multicolumn{1}{c|}{1156.054} &
		0.0201 \\ \hline
		FDCA Rules &
		\multicolumn{1}{c|}{\begin{tabular}[c]{@{}c@{}}{$\langle0,0,0,0,4,9,3,7\rangle$} \\ and {$\langle0,0,0,0,5,1,0,7\rangle$} \end{tabular}} &
		\multicolumn{1}{c|}{$\langle0,0,0,0,3,7,6,7\rangle$} &
		\multicolumn{1}{c|}{$\langle0,0,0,0,3,7,6,9\rangle$}  &
		{$\langle0,0,0,0,1,7,8,1\rangle$} \\ \hline
		Average Metric &
		\multicolumn{1}{c|}{0.521} &
		\multicolumn{1}{c|}{0.776} &
		\multicolumn{1}{c|}{1180.126} &
		0.1080 \\ \hline
		FDCA Rules &
		\multicolumn{1}{c|}{$\langle0,0,0,0,3,7,6,9\rangle$}  &
		\multicolumn{1}{c|}{$\langle0,0,0,0,3,9,4,9\rangle$} &
		\multicolumn{1}{c|}{$\langle0,0,0,0,3,9,4,9\rangle$}  &
		{$\langle0,0,0,0,4,9,3,1\rangle$}  \\ \hline
		Maximum participation &
		\multicolumn{1}{c|}{0.551} &
		\multicolumn{1}{c|}{0.0781} &
		\multicolumn{1}{c|}{1197.733} &
		0.0104 \\ \hline
		FDCA Rules &
		\multicolumn{1}{c|}{$\langle0,0,0,0,4,9,3,3\rangle$} &
		\multicolumn{1}{c|}{$\langle0,0,0,0,3,9,4,7\rangle$} &
		\multicolumn{1}{c|}{$\langle0,0,0,0,1,7,8,1\rangle$}  &
		{$\langle0,0,0,0,1,7,8,1\rangle$}  \\ \hline
	\end{tabular}
\end{table}

Table~\ref{FDCA Rules with Proposed Algorithm-1} and Table~\ref{FDCA Rules with Proposed Algorithm-2} showcases the best outcomes of clustering algorithms applied to four distinct datasets, presenting various evaluation metrics for each algorithm and their associated rules.
Through meticulous examination, it becomes evident that there exist notable discrepancies in the FDCA rules associated with different merging techniques. For instance, the FDCA rules linked to the average based merging technique encompass CAs such as $\langle0,0,0,0,3,9,4,7\rangle$ and $\langle0,0,0,0,6,7,3,9\rangle$, among others. Conversely, the maximum participation score based metric entails distinct FDCA rules like $\langle0,0,0,0,3,7,6,7\rangle$ and $\langle0,0,0,0,5,1,0,3\rangle$, among others. 

These diverse outcomes observed through FDCA rules underscore the algorithm's adaptability and its potential for further refinement and optimization. This comprehensive evaluation significantly contributes to the understanding of the algorithm's capabilities and lays the groundwork for potential avenues of future research and development in this domain.

 Table~\ref{Tab:Comparison Clustering-1} and Table~\ref{Tab:Comparison Clustering-2} shows the performance of our proposed algorithms with various existing clustering algorithms (K-means, Hierarchical, BIRCH, PAM, MeanShift, Binary CA) including the Sort G{\"o}del technique (Section~\ref{sec:SortGodel}) on the four datasets. We employ Silhouette score, Davies-Bouldin Index (DB), Calinski-Harabasz Index (CH), and Dunn Index to assess the clustering quality, indicating the algorithm's effectiveness in identifying meaningful data clusters. Our results reveal significant variations in clustering performance across the algorithms and datasets. This underscores the significance of choosing an algorithm that aligns with the unique attributes of the data under analysis. Moreover, integrating G{\"o}del numbers into the clustering procedure exhibited beneficial impacts on the overall quality of clustering.

\begin{sidewaystable}[!h]
	\centering
	\caption{Comparison of all clustering techniques based on evaluation metrics for Seeds and User Knowledge Modelling Dataset}
	\label{Tab:Comparison Clustering-1}
	\begin{tabular}{|ccccccccccc|}
		\hline
		\multicolumn{2}{|c|}{\multirow{3}{*}{\textbf{Algorithm}}} &
		\multicolumn{9}{c|}{\textbf{Dataset Name}}  \\ \hhline{~~---------} 
		\multicolumn{2}{|c|}{} &
		\multicolumn{1}{c|}{\multirow{2}{*}{\textbf{SCHEME}}}  &
		\multicolumn{4}{c|}{\textbf{Seeds}} &
		\multicolumn{4}{c|}{\textbf{User Knowledge Modelling}} \\ \hhline{~~~--------} 
		\multicolumn{2}{|c|}{} &
		\multicolumn{1}{c|}{} &
		\multicolumn{1}{c|}{\textbf{Silhouette}} &
		\multicolumn{1}{c|}{\textbf{DB}} &
		\multicolumn{1}{c|}{\textbf{CH}} &
		\multicolumn{1}{c|}{\textbf{Dunn Index}} &
		\multicolumn{1}{c|}{\textbf{Silhouette}} &
		\multicolumn{1}{c|}{\textbf{DB}} &
		\multicolumn{1}{c|}{\textbf{CH}} &
		\textbf{Dunn Index} \\ \hline
		\multicolumn{2}{|c|}{\multirow{2}{*}{\textbf{K-means}}} &
		\multicolumn{1}{c|}{\textbf{Normal}} &
		\multicolumn{1}{c|}{0.531} &
		\multicolumn{1}{c|}{0.659} &
		\multicolumn{1}{c|}{345.488} &
		\multicolumn{1}{c|}{0.0095} &
		\multicolumn{1}{c|}{0.265} &
		\multicolumn{1}{c|}{0.659} &
		\multicolumn{1}{c|}{345.488} &
		{0.1279} \\ \hhline{~~---------} 
		\multicolumn{2}{|c|}{} &
		\multicolumn{1}{c|}{\textbf{Godel number}} &
		\multicolumn{1}{c|}{0.952} &
		\multicolumn{1}{c|}{1.615} &
		\multicolumn{1}{c|}{150.607} &
		\multicolumn{1}{c|}{0.0072} &
		\multicolumn{1}{c|}{0.792} &
		\multicolumn{1}{c|}{0.326} &
		\multicolumn{1}{c|}{554.939} &
		{0.0016} \\ \hline
		\multicolumn{2}{|c|}{\multirow{2}{*}{\textbf{Hierarchical}}} &
		\multicolumn{1}{c|}{\textbf{Normal}} &
		\multicolumn{1}{c|}{0.495} &
		\multicolumn{1}{c|}{0.715} &
		\multicolumn{1}{c|}{269.977} &
		\multicolumn{1}{c|}{0.1050} &
		\multicolumn{1}{c|}{0.495} &
		\multicolumn{1}{c|}{0.722} &
		\multicolumn{1}{c|}{266.924} &
		{0.1050} \\ \hhline{~~---------} 
		\multicolumn{2}{|c|}{} &
		\multicolumn{1}{c|}{\textbf{Godel number}} &
		\multicolumn{1}{c|}{0.948} &
		\multicolumn{1}{c|}{0.137} &
		\multicolumn{1}{c|}{710.927} &
		\multicolumn{1}{c|}{0.0092} &
		\multicolumn{1}{c|}{0.839} &
		\multicolumn{1}{c|}{0.780} &
		\multicolumn{1}{c|}{648.566} &
		{0.0038} \\ \hline
		\multicolumn{2}{|c|}{\multirow{2}{*}{\textbf{BIRCH}}} &
		\multicolumn{1}{c|}{\textbf{Normal}} &
		\multicolumn{1}{c|}{0.346} &
		\multicolumn{1}{c|}{0.913} &
		\multicolumn{1}{c|}{129.199} &
		\multicolumn{1}{c|}{0.0035} &
		\multicolumn{1}{c|}{0.898} &
		\multicolumn{1}{c|}{0.108} &
		\multicolumn{1}{c|}{166480.013} &
		{0.0986} \\ \hhline{~~---------} 
		\multicolumn{2}{|c|}{} &
		\multicolumn{1}{c|}{\textbf{Godel number}} &
		\multicolumn{1}{c|}{0.978} &
		\multicolumn{1}{c|}{0.152} &
		\multicolumn{1}{c|}{3688.029} &
		\multicolumn{1}{c|}{0.2143} &
		\multicolumn{1}{c|}{0.965} &
		\multicolumn{1}{c|}{0.535} &
		\multicolumn{1}{c|}{360.694} &
		{0.1052} \\ \hline
		\multicolumn{2}{|c|}{\multirow{2}{*}{\textbf{PAM}}} &
		\multicolumn{1}{c|}{\textbf{Normal}} &
		\multicolumn{1}{c|}{0.435} &
		\multicolumn{1}{c|}{0.786} &
		\multicolumn{1}{c|}{197.331} &
		\multicolumn{1}{c|}{0.0053} &
		\multicolumn{1}{c|}{0.952} &
		\multicolumn{1}{c|}{0.051} &
		\multicolumn{1}{c|}{10442.09} &
		{0.1013} \\ \hhline{~~---------} 
		\multicolumn{2}{|c|}{} &
		\multicolumn{1}{c|}{\textbf{Godel number}} &
		\multicolumn{1}{c|}{0.766} &
		\multicolumn{1}{c|}{1.111} &
		\multicolumn{1}{c|}{53.132} &
		\multicolumn{1}{c|}{0.0093} &
		\multicolumn{1}{c|}{0.923} &
		\multicolumn{1}{c|}{0.824} &
		\multicolumn{1}{c|}{165.36} &
		{0.1463} \\ \hline
		\multicolumn{2}{|c|}{\multirow{2}{*}{\textbf{MeanShift}}} &
		\multicolumn{1}{c|}{\textbf{Normal}} &
		\multicolumn{1}{c|}{0.415} &
		\multicolumn{1}{c|}{0.823} &
		\multicolumn{1}{c|}{316.887} &
		\multicolumn{1}{c|}{0.0012} &
		\multicolumn{1}{c|}{0.952} &
		\multicolumn{1}{c|}{0.051} &
		\multicolumn{1}{c|}{10442.09} &
		{0.1013} \\ \hhline{~~---------} 
		\multicolumn{2}{|c|}{} &
		\multicolumn{1}{c|}{\textbf{Godel number}} &
		\multicolumn{1}{c|}{0.954} &
		\multicolumn{1}{c|}{0.132} &
		\multicolumn{1}{c|}{109806.136} &
		\multicolumn{1}{c|}{0.2155} &
		\multicolumn{1}{c|}{0.936} &
		\multicolumn{1}{c|}{0.157} &
		\multicolumn{1}{c|}{44715.827} &
		{0.0856} \\ \hline
		\multicolumn{2}{|c|}{\multirow{2}{*}{\textbf{Binary CA}}} &
		\multicolumn{1}{c|}{\textbf{Normal}} &
		\multicolumn{1}{c|}{0.502} &
		\multicolumn{1}{c|}{1.756} &
		\multicolumn{1}{c|}{123.892} &
		\multicolumn{1}{c|}{0.0010} &
		\multicolumn{1}{c|}{0.325} &
		\multicolumn{1}{c|}{0.616} &
		\multicolumn{1}{c|}{178.123} &
		{0.0029}\\ \hhline{~~---------} 
		\multicolumn{2}{|c|}{} &
		\multicolumn{1}{c|}{\textbf{Godel number}} &
		\multicolumn{1}{c|}{0.483} &
		\multicolumn{1}{c|}{0.716} &
		\multicolumn{1}{c|}{302.613} &
		\multicolumn{1}{c|}{0.0036} &
		\multicolumn{1}{c|}{0.441} &
		\multicolumn{1}{c|}{0.498} &
		\multicolumn{1}{c|}{303.260} &
		{0.0056} \\ \hline
		\multicolumn{3}{|c|}{\textbf{Sort Godel}} &
		\multicolumn{1}{c|}{0.493} &
		\multicolumn{1}{c|}{0.824} &
		\multicolumn{1}{c|}{318.944} &
		\multicolumn{1}{c|}{0.0072} &
		\multicolumn{1}{c|}{0.203} &
		\multicolumn{1}{c|}{0.993} &
		\multicolumn{1}{c|}{22910.758} &
		{0.0022} \\ \hline
		\multicolumn{1}{|c|}{\multirow{3}{*}{\textbf{\begin{tabular}[c]{@{}c@{}}Proposed \\ Algorithm \end{tabular}}}} &
		\multicolumn{2}{c|}{{\begin{tabular}[c]{@{}c@{}}\textbf{Silhouette} \\ \textbf{Score metric}\end{tabular}}} &
		\multicolumn{1}{c|}{\textbf{0.714}} &
		\multicolumn{1}{c|}{\textbf{10.113}} &
		\multicolumn{1}{c|}{\textbf{688.135}} &
		\multicolumn{1}{c|}{\textbf{0.0052}} &
		\multicolumn{1}{c|}{\textbf{0.783}} &
		\multicolumn{1}{c|}{\textbf{0.628}} &
		\multicolumn{1}{c|}{\textbf{249.447}} &
		\textbf{0.0080}  \\ \hhline{~----------} 
		\multicolumn{1}{|c|}{} &
		\multicolumn{2}{c|}{\textbf{Average Metric}} &
		\multicolumn{1}{c|}{\textbf{0.529}} &
		\multicolumn{1}{c|}{\textbf{10.231}} &
		\multicolumn{1}{c|}{\textbf{680.914}} &
		\multicolumn{1}{c|}{\textbf{0.0030}} &
		\multicolumn{1}{c|}{\textbf{0.756}} &
		\multicolumn{1}{c|}{\textbf{0.609}} &
		\multicolumn{1}{c|}{\textbf{247.675}} &
		\textbf{0.0082 }\\ \hhline{~----------} 
		\multicolumn{1}{|c|}{} &
		\multicolumn{2}{c|}{\textbf{Maximum participation}} &
		\multicolumn{1}{c|}{\textbf{0.795}} &
		\multicolumn{1}{c|}{\textbf{10.086}} &
		\multicolumn{1}{c|}{\textbf{685.351}} &
		\multicolumn{1}{c|}{\textbf{0.0032}} &
		\multicolumn{1}{c|}{\textbf{0.824}} &
		\multicolumn{1}{c|}{\textbf{0.808}} &
		\multicolumn{1}{c|}{\textbf{449.141}} &
		\textbf{0.0107} \\ \hline
	\end{tabular}
\end{sidewaystable}

\begin{sidewaystable}[]
	\centering
	\caption{Comparison of all clustering techniques based on evaluation metrics for Heart Failure Clinical and Buddy Move Dataset}
	\label{Tab:Comparison Clustering-2}
	\begin{tabular}{|ccccccccccc|}
		\hline
		\multicolumn{2}{|c|}{\multirow{3}{*}{\textbf{Algorithm}}} &
		\multicolumn{9}{c|}{\textbf{Dataset Name}}   \\ \hhline{~~---------}  
		\multicolumn{2}{|c|}{} &
		\multicolumn{1}{c|}{\multirow{2}{*}{\textbf{SCHEME}}} &
		\multicolumn{4}{c|}{\textbf{Heart Failure Clinical}} &
		\multicolumn{4}{c|}{\textbf{Buddy Move}} \\ \hhline{~~~--------}
		\multicolumn{2}{|c|}{} &
		\multicolumn{1}{c|}{} &
		\multicolumn{1}{c|}{\textbf{Silhouette}} &
		\multicolumn{1}{c|}{\textbf{DB}} &
		\multicolumn{1}{c|}{\textbf{CH}} &
		\multicolumn{1}{c|}{\textbf{Dunn Index}} &
		\multicolumn{1}{c|}{\textbf{Silhouette}} &
		\multicolumn{1}{c|}{\textbf{DB}} &
		\multicolumn{1}{c|}{\textbf{CH}} &
		\textbf{Dunn Index} \\ \hline
		\multicolumn{2}{|c|}{\multirow{2}{*}{\textbf{K-means}}} &
		\multicolumn{1}{c|}{\textbf{Normal}} &
		\multicolumn{1}{c|}{{0.583}} &
		\multicolumn{1}{c|}{{0.659}} &
		\multicolumn{1}{c|}{{266.924}} &
		\multicolumn{1}{c|}{{0.1285}} &
		\multicolumn{1}{c|}{{0.320}} &
		\multicolumn{1}{c|}{{1.336}} &
		\multicolumn{1}{c|}{{119.97}} &
		{0.045} \\ \hhline{~~---------} 
		\multicolumn{2}{|c|}{} &
		\multicolumn{1}{c|}{\textbf{Godel number}} &
		\multicolumn{1}{c|}{{0.617}} &
		\multicolumn{1}{c|}{{0.597}} &
		\multicolumn{1}{c|}{{495.442}} &
		\multicolumn{1}{c|}{{0.0070}} &
		\multicolumn{1}{c|}{{0.581}} &
		\multicolumn{1}{c|}{{15.861}} &
		\multicolumn{1}{c|}{{1.680}} &
		{0.0037} \\ \hline
		\multicolumn{2}{|c|}{\multirow{2}{*}{\textbf{Hierarchical}}} &
		\multicolumn{1}{c|}{\textbf{Normal}} &
		\multicolumn{1}{c|}{{0.679}} &
		\multicolumn{1}{c|}{{0.518}} &
		\multicolumn{1}{c|}{{411.563}} &
		\multicolumn{1}{c|}{{0.0052}} &
		\multicolumn{1}{c|}{{0.363}} &
		\multicolumn{1}{c|}{{1.018}} &
		\multicolumn{1}{c|}{{139.607}} &
		{0.1050}  \\ \hhline{~~---------} 
		\multicolumn{2}{|c|}{} &
		\multicolumn{1}{c|}{\textbf{Godel number}} &
		\multicolumn{1}{c|}{{0.635}} &
		\multicolumn{1}{c|}{{0.780}} &
		\multicolumn{1}{c|}{{425.585}} &
		\multicolumn{1}{c|}{{0.0064}} &
		\multicolumn{1}{c|}{{0.542}} &
		\multicolumn{1}{c|}{{0.515}} &
		\multicolumn{1}{c|}{{631.994}} &
		{0.0035} \\ \hline
		\multicolumn{2}{|c|}{\multirow{2}{*}{\textbf{BIRCH}}} &
		\multicolumn{1}{c|}{\textbf{Normal}} &
		\multicolumn{1}{c|}{{0.537}} &
		\multicolumn{1}{c|}{{0.611}} &
		\multicolumn{1}{c|}{{866.341}} &
		\multicolumn{1}{c|}{{0.0078}} &
		\multicolumn{1}{c|}{{0.290}} &
		\multicolumn{1}{c|}{{1.078}} &
		\multicolumn{1}{c|}{{122.622}} &
		{0.0001}\\ \hhline{~~---------} 
		\multicolumn{2}{|c|}{} &
		\multicolumn{1}{c|}{\textbf{Godel number}} &
		\multicolumn{1}{c|}{{0.993}} &
		\multicolumn{1}{c|}{{0.0}} &
		\multicolumn{1}{c|}{{$25 \times 10^{14}$}} &
		\multicolumn{1}{c|}{{0.3622}} &
		\multicolumn{1}{c|}{{0.954}} &
		\multicolumn{1}{c|}{{0.287}} &
		\multicolumn{1}{c|}{{1198.253}} &
		{0.0042} \\ \hline
		\multicolumn{2}{|c|}{\multirow{2}{*}{\textbf{PAM}}} &
		\multicolumn{1}{c|}{\textbf{Normal}} &
		\multicolumn{1}{c|}{{0.535}} &
		\multicolumn{1}{c|}{{0.625}} &
		\multicolumn{1}{c|}{{879.158}} &
		\multicolumn{1}{c|}{{0.0056}} &
		\multicolumn{1}{c|}{{0.167}} &
		\multicolumn{1}{c|}{{1.914}} &
		\multicolumn{1}{c|}{{77.084}} &
		{0.0030} \\ \hhline{~~---------}
		\multicolumn{2}{|c|}{} &
		\multicolumn{1}{c|}{\textbf{Godel number}} &
		\multicolumn{1}{c|}{{0.668}} &
		\multicolumn{1}{c|}{{1.653}} &
		\multicolumn{1}{c|}{{147.029}} &
		\multicolumn{1}{c|}{{0.0078}} &
		\multicolumn{1}{c|}{{0.741}} &
		\multicolumn{1}{c|}{{0.611}} &
		\multicolumn{1}{c|}{{245.793}} &
		{0.0040} \\ \hline
		\multicolumn{2}{|c|}{\multirow{2}{*}{\textbf{MeanShift}}} &
		\multicolumn{1}{c|}{\textbf{Normal}} &
		\multicolumn{1}{c|}{{0.616}} &
		\multicolumn{1}{c|}{{0.504}} &
		\multicolumn{1}{c|}{{837.169}} &
		\multicolumn{1}{c|}{{0.0062}} &
		\multicolumn{1}{c|}{{0.293}} &
		\multicolumn{1}{c|}{{42.296}} &
		\multicolumn{1}{c|}{{95.265}} &
		{0.0001}  \\ \hhline{~~---------}
		\multicolumn{2}{|c|}{} &
		\multicolumn{1}{c|}{\textbf{Godel number}} &
		\multicolumn{1}{c|}{{0.997}} &
		\multicolumn{1}{c|}{{0.0}} &
		\multicolumn{1}{c|}{{$28 \times 10^{14}$}} &
		\multicolumn{1}{c|}{{0.3655}} &
		\multicolumn{1}{c|}{{0.774}} &
		\multicolumn{1}{c|}{{0.21}} &
		\multicolumn{1}{c|}{{1617.586}} &
		{0.0048} \\ \hline
		\multicolumn{2}{|c|}{\multirow{2}{*}{\textbf{Binary CA}}} &
		\multicolumn{1}{c|}{\textbf{Normal}} &
		\multicolumn{1}{c|}{{0.612}} &
		\multicolumn{1}{c|}{{1.265}} &
		\multicolumn{1}{c|}{{625.221}} &
		\multicolumn{1}{c|}{{0.0024}} &
		\multicolumn{1}{c|}{{0.367}} &
		\multicolumn{1}{c|}{{12.543}} &
		\multicolumn{1}{c|}{{59.108}} &
		{0.0041} \\ \hhline{~~---------}
		\multicolumn{2}{|c|}{} &
		\multicolumn{1}{c|}{\textbf{Godel number}} &
		\multicolumn{1}{c|}{{0.609}} &
		\multicolumn{1}{c|}{{0.895}} &
		\multicolumn{1}{c|}{{379.256}} &
		\multicolumn{1}{c|}{{0.0035}} &
		\multicolumn{1}{c|}{{0.403}} &
		\multicolumn{1}{c|}{{1.118}} &
		\multicolumn{1}{c|}{{73.442}} &
		{0.0063} \\ \hline
		\multicolumn{3}{|c|}{\textbf{Sort Godel}} &
		\multicolumn{1}{c|}{{0.383}} &
		\multicolumn{1}{c|}{{0.777}} &
		\multicolumn{1}{c|}{{58.740}} &
		\multicolumn{1}{c|}{{0.0035}} &
		\multicolumn{1}{c|}{{0.233}} &
		\multicolumn{1}{c|}{{0.962}} &
		\multicolumn{1}{c|}{{51.456}} &
		{0.0080} \\ \hline
		\multicolumn{1}{|c|}{\multirow{3}{*}{\textbf{\begin{tabular}[c]{@{}c@{}}Proposed \\ Algorithm\end{tabular}}}} &
		\multicolumn{2}{c|}{\textbf{\begin{tabular}[c]{@{}c@{}}Silhouette Score\\ metric\end{tabular}}} &
		\multicolumn{1}{c|}{\textbf{0.619}} &
		\multicolumn{1}{c|}{\textbf{1.571}} &
		\multicolumn{1}{c|}{\textbf{837.147}} &
		\multicolumn{1}{c|}{\textbf{0.0107}} &
		\multicolumn{1}{c|}{\textbf{0.536}} &
		\multicolumn{1}{c|}{\textbf{0.782}} &
		\multicolumn{1}{c|}{\textbf{1156.054}} &
		\textbf{0.0201} \\ \hhline{~----------} 
		\multicolumn{1}{|c|}{} &
		\multicolumn{2}{c|}{\textbf{Average Metric}} &
		\multicolumn{1}{c|}{\textbf{0.587}} &
		\multicolumn{1}{c|}{\textbf{0.651}} &
		\multicolumn{1}{c|}{\textbf{248.191}} &
		\multicolumn{1}{c|}{\textbf{0.0107}} &
		\multicolumn{1}{c|}{\textbf{0.521}} &
		\multicolumn{1}{c|}{\textbf{0.776}} &
		\multicolumn{1}{c|}{\textbf{1180.126}} &
		\textbf{0.1080} \\ \hhline{~----------} 
		\multicolumn{1}{|c|}{} &
		\multicolumn{2}{c|}{\textbf{Maximum participation}} &
		\multicolumn{1}{c|}{\textbf{0.728}} &
		\multicolumn{1}{c|}{\textbf{1.604}} &
		\multicolumn{1}{c|}{\textbf{822.47}} &
		\multicolumn{1}{c|}{\textbf{0.0108}} &
		\multicolumn{1}{c|}{\textbf{0.551}} &
		\multicolumn{1}{c|}{\textbf{0.0781}} &
		\multicolumn{1}{c|}{\textbf{1197.733}} &
		\textbf{0.0104} \\ \hline
	\end{tabular}
\end{sidewaystable}

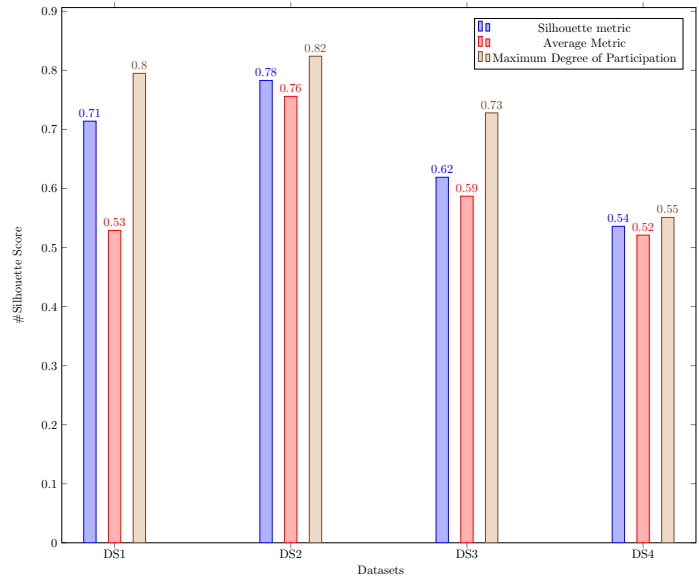
\begin{figure}[!h]
	\centering
	\resizebox{0.7\textwidth}{!}{
		\pgfplotsset{width=1.5\textwidth} 
		\begin{tikzpicture}
			\begin{axis}
				[symbolic x coords={DS1, DS2, DS3, DS4}, xtick=data, xticklabel style={text height=1ex}, 
				ymin=0, ybar=10pt, nodes near coords, nodes near coords align={vertical}, 
				ylabel={\#Silhouette Score}, xlabel={\ Datasets}] 
				\addplot coordinates 
				{(DS1,0.714) (DS2,0.783) (DS3,0.619) (DS4,0.536) };
				\addplot coordinates 
				{(DS1, 0.529) (DS2, 0.756) (DS3, 0.587) (DS4, 0.521) };
				\addplot coordinates 
				{(DS1, 0.795) (DS2, 0.824) (DS3, 0.728) (DS4, 0.551) };
				\legend{Silhouette metric, Average Metric, Maximum Degree of Participation} 
			\end{axis}
	\end{tikzpicture}}
	\caption{Comparison of Decimal FDCA Cluster Merging Techniques based on Silhouette Score}
	\label{Plot-1}
\end{figure}

Figure~\ref{Plot-1} presents the comparative analysis of decimal FDCA cluster merging techniques across the four distinct datasets. The Silhouette score, depicted on the y-axis, serves as a pivotal metric for evaluating clustering quality. This score quantifies the similarity of each object to its assigned cluster in contrast to other clusters.
Three primary merging techniques are scrutinized in the analysis: Silhouette score based metric, average based metric, and maximum degree of participation. For each dataset, three bars represent the Silhouette score attained by employing these techniques. The height of each bar correlates directly with the Silhouette score value, providing a clear visual representation of the clustering performance.
A higher Silhouette score signifies superior clustering quality, indicating that objects within each cluster exhibit greater similarity while being distinct from objects in other clusters. We can see that, the maximum participation score based metric performs better for merging into meaningful clusters.

\begin{figure}[!h]
	\centering
	\resizebox{0.7\textwidth}{!}{
		\pgfplotsset{width=1.5\textwidth} 
		\begin{tikzpicture}
			\begin{axis}
				[symbolic x coords={DS1, DS2, DS3, DS4}, xtick=data, xticklabel style={text height=1ex}, ymin=0, ybar=10pt, nodes near coords, nodes near coords align={vertical}, ylabel={\#Silhouette Score}, xlabel={\ Datasets}] 
				\addplot coordinates 
				{(DS1,0.493) (DS2,0.203) (DS3,0.383) (DS4,0.233) };
				\addplot coordinates 
				{(DS1, 0.531) (DS2, 0.265) (DS3, 0.583) (DS4, 0.320)};
				\addplot coordinates 
				{(DS1, 0.795) (DS2, 0.824) (DS3, 0.728) (DS4, 0.551) };
				\addplot coordinates 
				{(DS1, 0.952) (DS2, 0.792) (DS3, 0.617) (DS4, 0.581)}; 
				
				\legend
				{Sort G{\"o}del, K-Means, FDCA with maximum degree of participation, K-Means with G{\"o}del} 
			\end{axis}	
	\end{tikzpicture}}
	\caption{Comparison of Techniques based on Silhouette Score}
	\label{Plot-2}
\end{figure}
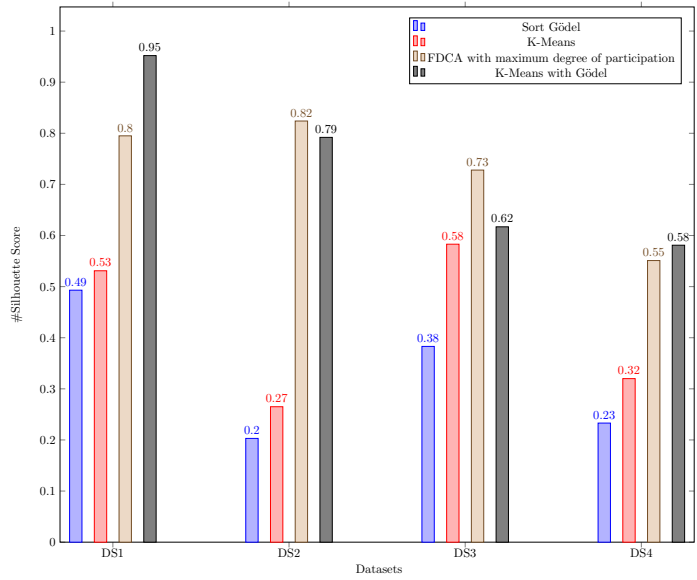

Similarly, the bar graph shown in Figure~\ref{Plot-2} presents a comparative analysis of various clustering techniques based on their Silhouette score across four distinct datasets. The analysis evaluates four different techniques: Sort G{\"o}del, K-Means, FDCA with maximum degree of participation, and K-Means with G{\"o}del. Each technique's performance is represented by a group of bars, with each bar corresponding to a specific dataset displayed along the x-axis. This graph clearly shows that, our decimal FDCA based clustering that uses G{\"o}del number based encoding outperforms the quality of all existing state-of-the-art techniques. Moreover, it also indicates that including G{\"o}del number based encoding in the existing state-of-the-art technique can improve the cluster quality of those algorithms. 

\section{Conclusion}
In this paper, we propose G{\"o}del number-based encoding as a superior alternative to Hash-based encoding and frequency-based encoding techniques. This encoding method preserves the essential properties of features, resulting in more meaningful clusters. Over the G{\"o}del encoded dataset, we apply decimal first degree cellular automata for clustering. Our algorithm works in three stages and considers three metrics as options for achieving the desired number of clusters. Apart from the Silhouette score, the other proposed metrics are based on average cluster distance and maximum degree of participation.

We have selected $36$ rules through theoretical analysis over chaotic property and the cycle structure of the  decimal FDCA rules. In our experiments, we have tested our algorithm with these rules across four datasets. We evaluate cluster quality using metrics such as the Silhouette score, DB score, CH score, and Dunn index, compare our approach with existing data mining algorithms. It is observed that, the maximum degree of participation based metric for merging gives much better result than the other metrics and is at par with the state-of-the-art algorithm. Moreover, we also find that applying G{\"o}del number-based encoding before utilizing existing algorithms like K-Means or Hierarchical clustering consistently yields better results compared to other encoding techniques.

Our research highlights the potential of decimal cellular automata in clustering applications across a wide range of datasets. The inherent ability of CA to capture complex patterns and self-organize presents it as a promising alternative to traditional clustering algorithms. Overall, our findings demonstrate the effectiveness of G{\"o}del number-based encoding and the potential of cellular automata in clustering applications, offering comparable results to traditional clustering algorithms like K-Means, Hierarchical, and BIRCH.
However, further optimization of CA rules and parameter tuning could significantly enhance its performance.  Also, exhaustive testing with all decimal rules could potentially yield even better results. To optimize computational efficiency, we propose splitting the data frame into subsets and running each split independently, allowing for multi-threaded parallel processing to reduce time complexity.

\textbf{Acknowledgments:}
The authors are grateful to Prof. Mihir Chakraborty for introducing us to the beautiful world of logic and teaching G{\"o}del numbering, and to Prof. Sukanta Das for his valuable advice and discussions on the design and implementation of the scheme.

\section*{Declarations}

\textbf{Ethical approval:} This is not applicable. \\

\textbf{Competing interests:} The authors have no relevant financial or nonfinancial interests to disclose.\\

\textbf{Author contributions:} \\Vicky Vikrant- Validation, Formal analysis, Investigation, Writing - Original Draft, Visualization. \\
Narodia Parth P- Software, Validation, Data Curation. \\
Kamalika Bhattacharjee- Conceptualization, Methodology, Writing - Review and Editing, Supervision, Funding acquisition.\\

\textbf{Funding:} 
This work is partially supported by Start-up Research Grant (File number: SRG/2022/002098), SERB, Department of Science \& Technology, Government of India, and NIT, Tiruchirappalli SEED Grant. \\

\textbf{Data availability:} No Data associated in the manuscript.


\bibliography{References.bib}

\end{document}